\documentclass[aps,prd,nofootinbib]{revtex4}

 \usepackage[utf8]{inputenc}
\usepackage{amsmath,amssymb,graphicx}
\usepackage{tabularx}
\usepackage{enumitem}
\usepackage{float}

\usepackage{tikz}

\usetikzlibrary{snakes}
\usetikzlibrary{calc,positioning,shapes.geometric,shapes.symbols,shapes.misc, intersections, shadows, arrows, arrows.meta}
\usetikzlibrary{backgrounds}

\newcommand{\be}{\begin{equation}}
\newcommand{\ee}{\end{equation}}
\newcommand{\bea}{\begin{eqnarray}}
\newcommand{\eea}{\end{eqnarray}}

\def\code#1{\texttt{#1}}

\begin{document}
\author{Alessandro Casalino}
\email{alessandro.casalino@unitn.it}
\author{Massimiliano Rinaldi}
\email{massimiliano.rinaldi@unitn.it}
\affiliation{Dipartimento di Fisica, Universit\`{a} degli Studi di Trento,\\Via Sommarive 14, I-38123 Povo (TN), Italy}
\affiliation{Trento Institute for Fundamental Physics and Applications (TIFPA)-INFN,\\Via Sommarive 14, I-38123 Povo (TN), Italy}

\title{Testing Horndeski gravity as dark matter with  \code{hi\_class}}

\begin{abstract}
\noindent We consider a model of dark matter fluid based on a sector of Horndeski gravity. The model is very successful, at the background level, in reproducing the evolution of the Universe from early times to today. However, at the perturbative level the model fails. To show this, we use the code \code{hi\_class} and we compute the matter power spectrum and the cosmic microwave background spectrum. Our results confirm, in a new and independent way, that this sector of Horndeski gravity is not viable to describe dark matter, in agreement with the recent constraints coming from the measurement of the speed of gravitational waves obtained from the observation of the neutron star merger event GW170817. 
\end{abstract}

\maketitle

\section{Introduction}

\noindent The recent observation of the neutron star merger event GW170817 \cite{TheLIGOScientific:2017qsa} has put a very strong constraint on the speed of gravitational waves $c_{T}$, greatly helping the effort of constraining the huge set of Modified General Relativity models (see other constraints analysis, e.g. \cite{Hou:2017cjy}-\cite{Bhattacharya:2016naa}). Indeed this constraint has invalidated, or severely constrained, many models of modified gravity, as anticipated already in  \cite{Lombriser:2015sxa}-\cite{Lombriser:2016yzn}. In particular, the Horndeski model \cite{Horndeski:1974wa} seems no longer admit terms with $G_5(\phi,X)$ or $G_4$ functions with a $X$ dependence \cite{Creminelli:2017sry}-\cite{Baker:2017hug}. Nevertheless, the value of $c_{T}$ has been measured by a single merger event. Therefore, an independent way to rule out a theory can be very useful, at least until several measurements of $c_{T}$ will be available. In this paper we focus on a particular sector of Horndeski gravity that appears to be a very promising candidate for cosmological dark matter (and maybe also for the anomalous rotation curves of spiral galaxies) at the level of background equations of motion \cite{Rinaldi_2016}. The core of the model is the non-minimal coupling of the Einstein tensor to the kinetic term of the scalar field in the Lagrangian (also proposed long ago by Amendola in \cite{amendola}), which shows interesting properties not only at cosmological level. Indeed, since the discovery of a simple black hole analytical solution in \cite{maxBH}, this sector of the theory yielded more general viable black hole solutions and realistic neutron stars \cite{NSHorn}-\cite{NSHorn2}, see also \cite{babichev}. 

It is well-known that the non-minimal coupling between Einstein tensor and scalar kinetic term leads to a different speed of the gravitational waves than light. Thus, in principle, there is no longer need to study these models since they appear to be ruled out by the observation of the event GW170817. However, it is a good practice to find more than one motivation to discard a theory. In fact, one might argue that the speed of gravitational waves that has been measured corresponds to a narrow range of wavelengths that came from a single event in the late Universe. Since we are dealing with a cosmological model that goes back to inflation, we should investigate other ways to prove or disprove the model. In addition,  this investigation might shed further light on why Horndeski gravity is fundamentally ruled out by Nature.

In this paper, we use a recent and powerful software, \code{hi\_class}, specifically designed by Zumalacàrregui et al.\ \cite{Zumalacarregui:2016pph} to calculate the power spectra in Horndeski gravity. Our aim is to show that the theory disagrees with the observations of the Cosmic Microwave Background (CMB) and also of the matter power spectrum, independently of the bound on $c_{T}$ established by the observation of the event GW170817. In particular, we show that the disagreement persists even when we adapt the parameters of the model to meet the experimental constraints on $c_{T}$. Even more, we find that the non-minimally coupled scalar field is always incompatible with observations, even in the case it does not contribute to the dark matter content of the Universe. 
Similar results, using a Gleyzes-Langlois-Piazza-Vernizzi action that includes also our model, were recently found in \cite{Diez-Tejedor:2018fue}.

The structure of the paper is the following. In the next section we recall the essentials of the model that was studied in \cite{Rinaldi_2016}. In section \ref{sec:aCS_background} we study the background solutions of the model numerically for different values of the parameters in the action, using the software \code{hi\_class}. In section \ref{sec:stability} we analyse the stability of the perturbation functions, arguing that the speed of scalar perturbations $c_s^2$ can become negative during certain epochs of the evolution of the Universe. The core of the paper is in section \ref{sec:perturbations}, where we study the perturbations in detail, by computing the CMB and the matter power spectrum. We will analyse the instabilities and explain their origin. In the end, in section \ref{sec:gravity_waves}, we will find in which range the  parameters of the theory are compatible with the bounds on $c_{T}$ from the event GW170817.

\section{The model}\label{sec2}

\noindent Before considering the specific model studied in \cite{Rinaldi_2016}, we recall some basic facts about Horndeski gravity. The total Lagrangian reads
\begin{equation}
S[g_{\mu \nu},\phi]=\frac{1}{8 \pi G}\int d^4x \sqrt{-g} \sum_{i=2}^5 \mathcal{L}_i[g_{\mu \nu},\phi] + S_{M}[g_{\mu \nu}]\,,
\label{eq:CH(MGR)_horndeski_action}
\end{equation}
where the  $\mathcal{L}_i$ are defined by
\begin{align}
\mathcal{L}_2 &= G_2 [\phi, X], \label{eq:CH(MGR)_horndeski_action_L_2}\\
\mathcal{L}_3 &= -G_3 [\phi, X] \Box \phi, \label{eq:CH(MGR)_horndeski_action_L_3}\\
\mathcal{L}_4 &= G_4 [\phi, X] R + G_{4X} [\phi, X]\left[(\Box \phi)^2-\phi_{;\mu \nu}\phi^{;\mu \nu}\right],\label{eq:CH(MGR)_horndeski_action_L_4}\\
\mathcal{L}_5 &= G_5 [\phi, X] G_{\mu \nu}\phi^{;\mu \nu}-\frac{1}{6}G_{5X}[\phi, X]\left[(\Box \phi)^3-3(\Box \phi)\phi_{;\mu \nu}\phi^{;\mu \nu}+2\phi_{;\mu}^{\phantom{;\mu}\nu}\phi_{;\nu}^{\phantom{;\nu}\alpha}\phi_{;\alpha}^{\phantom{;\alpha}\mu}\right].\label{eq:CH(MGR)_horndeski_action_L_5}
\end{align}
The $G_i[\phi,X]$ are four arbitrary functions of the scalar field $\phi$, which represents a new degree of freedom, and $X=- \partial_\mu \phi \partial^{\mu} \phi/2$. Here, $\Box=\nabla_{\mu} \nabla^{\mu}$, and the subscript $\phi$ or $X$ on the $G_i$ functions represents respectively the derivative with respect to $\phi$ and $X$, e.g. $G_{iX}\equiv\partial G_i/\partial X$ and $G_{i\phi}\equiv\partial G_i/\partial \phi$ (and similarly for higher order derivatives).

The action of the model studied in \cite{Rinaldi_2016} is a subclass of the above theory and reads
\begin{equation}
S[g_{\mu \nu}, \phi]=\int d^4x \sqrt{-g} \left[\kappa (R-2\Lambda) - \frac{1}{2} (\alpha g_{\mu \nu} - \xi G_{\mu \nu}) \nabla^{\mu} \phi \nabla^{\nu} \phi\right] + S_m[g_{\mu \nu}],
\label{eq:CH(aCS)_action}
\end{equation}
where $\alpha$ and $\xi$ are the two parameters of the theory and $\Lambda$ corresponds to the observed cosmological constant. In the original model, the matter action $S_m[g_{\mu \nu}]$ describes only radiation and baryonic matter. Later on we will look at the case when it includes also some of the dark matter in the Universe. For what follows, it is important to stress that the theory is symmetric under constant shifts of the scalar field $\phi$.
The above action che be found by setting 
\bea\label{Gii}
G_2 [\phi, X] =-\Lambda+{1\over 2\kappa} \alpha X\,,\quad G_3 [\phi, X] = 0\,,\quad G_4 [\phi, X] = \frac{1}{2}\,,\quad G_5 [\phi, X] = \frac{1}{4\kappa}\xi \phi\,,
\eea
in \eqref{eq:CH(aCS)_action} and integrating by parts.

The modified Einstein equations read
\begin{equation}
G_{\mu \nu} + \Lambda g_{\mu \nu} + H_{\mu \nu} = {1\over 2\kappa} T_{\mu \nu}
\label{eq:CH(aCS)_eom}
\end{equation}
where  $H_{\mu \nu}$ is given by
\begin{eqnarray}
H_{\mu\nu}&=&-\frac{\alpha}{2\kappa}\Bigg[\nabla_{\mu}\phi\nabla_{\nu}\phi-\frac{1}{2}g_{\mu\nu}\nabla_{\lambda}\phi\nabla^{\lambda}\phi\Bigg]\nonumber-\frac{\xi}{2\kappa}\Bigg[\frac{1}{2}\nabla_{\mu}\phi\nabla_{\nu}\phi R-2\nabla_{\lambda}\phi\nabla_{(\mu}\phi R_{\nu)}^{\lambda}\nonumber\\
&-&\nabla^{\lambda}\phi\nabla^{\rho}\phi R_{\mu\lambda\nu\rho}-(\nabla_{\mu}\nabla^{\lambda}\phi)(\nabla_{\nu}\nabla_{\lambda}\phi)+\frac{1}{2}g_{\mu\nu}(\nabla^{\lambda}\nabla^{\rho}\phi)(\nabla_{\lambda}\nabla_{\rho}\phi)-\frac{1}{2}g_{\mu\nu}(\square\phi)^{2}\nonumber\\
&&+(\nabla_{\mu}\nabla_{\nu}\phi)\square\phi+\frac{1}{2}G_{\mu\nu}(\nabla\phi)^{2}+g_{\mu\nu}\nabla_{\lambda}\phi\nabla_{\rho}\phi R^{\lambda\rho}\Bigg] \,,
\end{eqnarray}
where $\kappa=(16\pi G)^{-1}$.

If we choose the standard flat Robertson-Walker metric 
\bea
ds^{2}=-dt^{2}+a(t)^{2}\delta_{ij}dx^{i}dx^{j}\,,
\eea
and we define $\psi\equiv \dot\phi$, the Friedmann equations take the  form
\bea
H^{2}&=&  {4\Lambda \kappa+\alpha\psi^{2}+2\rho_{r}+2\rho_{m}\over 3(4\kappa-3\xi\psi^2)}\,,\\\nonumber
\dot H&=& -{3\rho_{m}+4\rho_{r}\over 3(4\kappa-3\xi\psi^{2})}+{\psi\dot\psi(\alpha+9\xi H^{2})\over 3H(4\kappa-3\xi\psi^{2})}  \,,
\eea
where $H=\dot a/a$, $\rho_{r}$ and $\rho_{m}$ are the energy densities of radiation and non-relativistic fluid respectively, which satisfy the usual equations
\bea
\dot\rho_{m}=-3H\rho_{m}\,,\qquad \dot\rho_{r}=-4H\rho_{r}\,.
\eea
The Klein-Gordon equation for $\phi$ can be solved analytically to yield the relation
\begin{equation}
\psi = \frac{q}{a^2(\alpha + 3 \xi H^2)},
\label{eq:CH(aCS)_field_solution_analytical}
\end{equation}
where $q$ is an integration constant. We associate to the scalar field the fractional density
\begin{equation}
\Omega_\phi = \frac{q^2 (\alpha + 9\xi H^2)}{12\kappa H^2 a^6 (\alpha + 3\xi H^2)^2}\,,
\label{eq:CH(aCS)_fractional_density_analytical}
\end{equation}
that will be useful to  check the accuracy of the results found with \code{hi\_class}. The sum of all densities must always satisfy the constraint
\begin{equation}
\sum_{ i} \Omega_i = 1,
\label{eq:CH(aCS)_sum_rule}
\end{equation}
where $\Omega_{i}$ is the density of the $i^{\rm th}$ species that is baryonic matter, radiation and dark energy. In the case $\phi$ is not mimicking dark matter, or it is just one component of dark matter, we will add its density parameter to the sum. Finally, we define the new parameter
\begin{equation}
\beta = \xi \Lambda.
\label{eq:CH(aCS)_beta}
\end{equation}
which will be central in the numerical calculations below.

Before going further, we stress that the free parameters  are $\alpha$ and $\xi$, together with the initial values of $\psi$.  However, $\alpha$ can be eliminated through the substitution $\phi^2 \rightarrow \alpha \phi^2$ (the scalar field appears always quadratically in the equations of motion). Thus, in the following we will set $\alpha=0,\pm 1$ only. Therefore, the number of free parameters reduces to one, plus the initial condition on $\psi$.

Numerically, in order to use \code{hi\_class} we need to choose a parameter to apply a shooting method, which ensures that  (\ref{eq:CH(aCS)_sum_rule}) is satisfied. In most cases, we will consider $\psi$ (or equivalently $q$) as the shooting parameter to vary, and therefore its value will be chosen by the program.

The code  \code{hi\_class} solves the background equations of motion in terms of $5$ functions, which in our model are defined as follows
\begin{align}
M_*^2 &=  1- {1\over 4\kappa} \xi  \psi^2 \label{eq:CH(aCS)_horndeski_M*^2}\\
\alpha_M &= - \frac{2 \xi  \psi \dot{\psi} }{H^2 \left(2\kappa-\frac{\xi \psi^2}{2} \right)}\,,\label{eq:CH(aCS)_horndeski_alpha_M*^2_running}\\
\alpha_K &= \frac{\psi^2 \left(\alpha +3 \xi  H^2 \right)}{H^2 \left(2\kappa -\frac{\xi \psi^2}{2}\right)}\,,\label{eq:CH(aCS)_horndeski_alpha_kineticity}\\
\alpha_B &=\frac{2 \xi  \psi^2}{2\kappa-\frac{\xi \psi^2}{2}}\,,\label{eq:CH(aCS)_horndeski_alpha_braiding}\\
\alpha_T &=\frac{\xi  \psi^2}{2\kappa-\frac{\xi \psi^2}{2}}=\frac{1}{2}\alpha_B\label{eq:CH(aCS)_horndeski_alpha_tensor_excess}.
\end{align}
These are in turn used for the parametrisation of the linear perturbations, as shown in \cite{Zumalacarregui:2016pph}.
In order to use \code{hi\_class} we also have to specify the form of the energy density and pressure of the field. We find

\begin{align}
\rho_\phi &= \frac{9}{2} \xi  H^2 \psi^2+\frac{1}{2} \alpha  \psi^2 \label{eq:CH(aCS)_horndeski_rho}\\
P_\phi &= \frac{ a \left(\xi  H^2 +\alpha \right) \psi^3 - 2 \xi  \dot{H}\psi^3 - 4 \xi  H \psi^2 \dot{\psi}}{2 a \psi}, \label{eq:CH(aCS)_horndeski_pressure}
\end{align}
and the equation of state (EoS) parameter 
\begin{equation}
\omega_\phi = \frac{P_\phi}{\rho_\phi} = \frac{a \left(\xi  H^2+\alpha \right)\psi^3-2 \xi  \dot{H}\psi^3-4 \xi  H \psi^2 \dot{\psi}}{a \left(9 \xi  H^2+\alpha \right)\psi^3}.
\end{equation}

In general, the implementation of a model in \code{hi\_class} with the functions \eqref{eq:CH(aCS)_horndeski_M*^2}-\eqref{eq:CH(aCS)_horndeski_alpha_tensor_excess} requires a parametrisation depending only on the background quantities $a$, $H$ and $\dot{H}$. In our case, this is possible thanks to Eq.\  (\ref{eq:CH(aCS)_field_solution_analytical}), that relates $\psi$ to $a$, $H$ and $\dot{H}$. In Appendix \ref{sec:numerical_evaluation_alphas} we show in detail how to approach the problem numerically in \code{hi\_class}.

In addition to the implementation with the $\alpha$ functions \eqref{eq:CH(aCS)_horndeski_M*^2}-\eqref{eq:CH(aCS)_horndeski_alpha_tensor_excess}, \code{hi\_class} offers the possibility to evolve the model numerically using  directly Eqs.\ \eqref{Gii} as input functions. As already mentioned, the first implementation uses only functions of the background quantities $a$, $H$ and $\dot H$, and there is no need to find the time evolution of the scalar field $\phi$. The second implementation requires the definition of the action of the model by specifying the $G_i$ functions, and solves the Klein-Gordon equation in order to find the evolution of the scalar field. The first possibility is realised thanks to the public version of \code{hi\_class} \cite{Zumalacarregui:2016pph}, while the second is only possibile with the developer's one. 
\section{Background evolution}
\label{sec:aCS_background}

\begin{figure}[H]
  \begin{center}
    \includegraphics[width=1.0\textwidth]{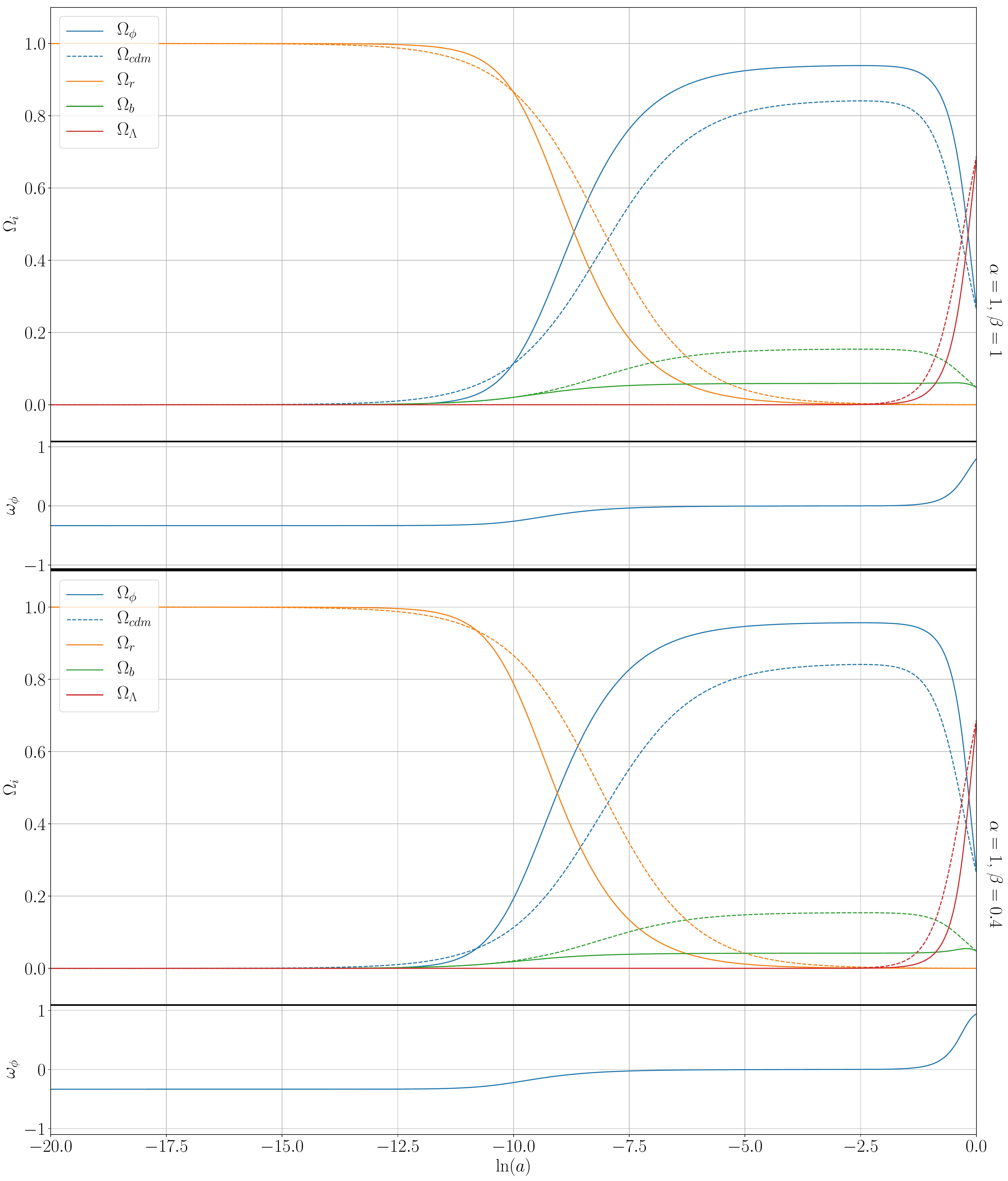}
  \end{center}
  \caption{Plot of the fractional densities and of $\omega_\phi$  as functions of the e-folding number $N$, computed with \code{hi\_class}. Here, $\alpha=1$ and $\beta$ takes two positive values. The dashed lines corresponds to  the standard $\Lambda$CDM model.}  \label{fig:CH(aCS)_rinaldi_background}
\end{figure}

\begin{figure}[H]
  \begin{center}
    \includegraphics[width=1.0\textwidth]{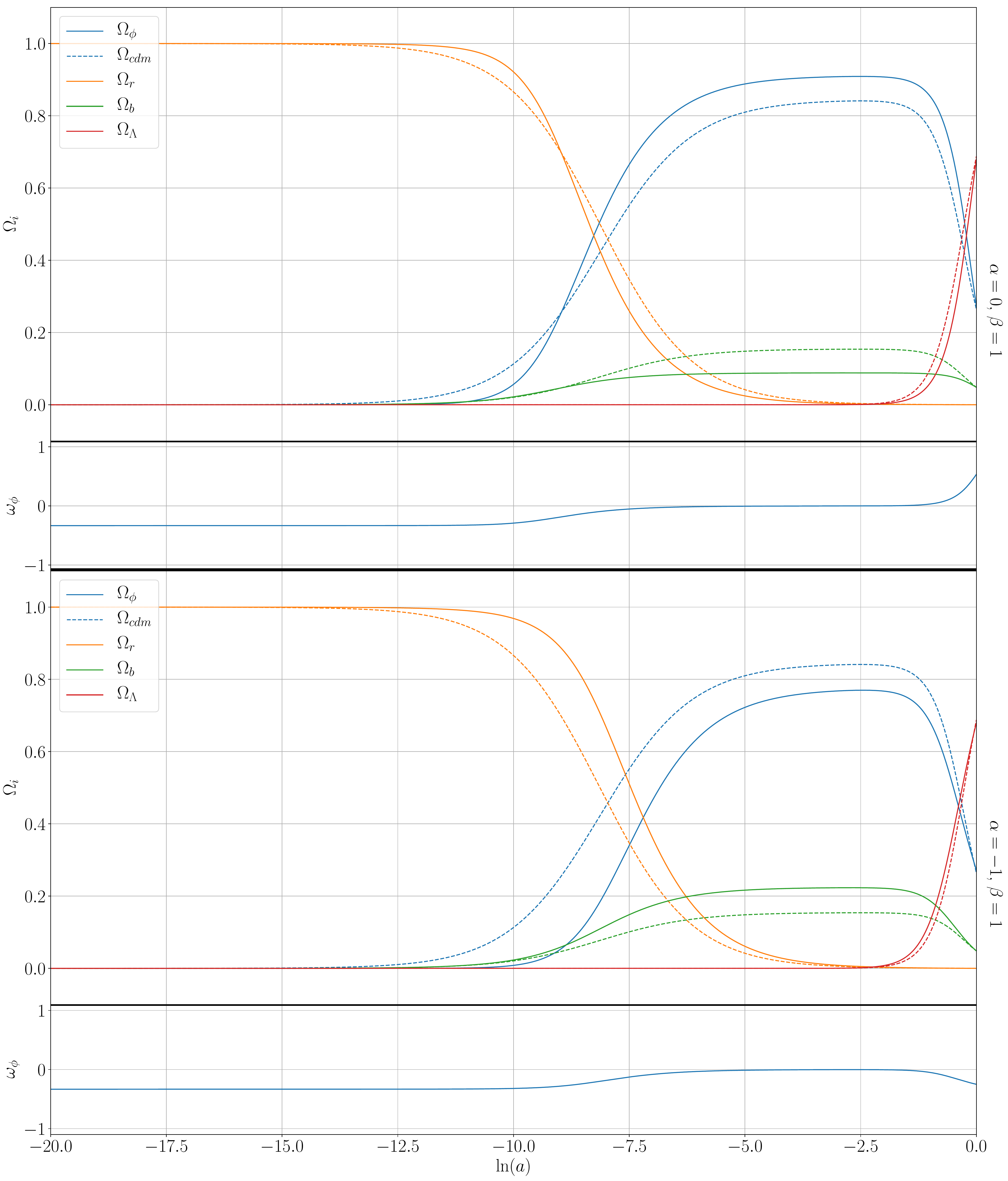}
  \end{center}
  \caption{Plot of the fractional densities and of $\omega_\phi$  as functions of the e-folding number $N$, computed with \code{hi\_class}. Here, $\alpha=1$ and $\beta$ takes two positive values. The dashed lines corresponds to  the standard $\Lambda$CDM model.}
  \label{fig:CH(aCS)_rinaldi_background2}
\end{figure}

\noindent Let us briefly analyse the background solutions of the model, using the range of parameters adopted in \cite{Rinaldi_2016}. In Figs.\ \ref{fig:CH(aCS)_rinaldi_background} and \ref{fig:CH(aCS)_rinaldi_background2} we show the fractional densities of the cosmological components and the scalar field equation of state parameter $\omega_\phi$.
We note that for $\beta\sim 1$ the scalar field behaves as the dark matter fluid in $\Lambda$CDM. This is evident also in Figs.\ \ref{fig:CH(aCS)_rinaldi_background} and \ref{fig:CH(aCS)_rinaldi_background2}, where we see that $\omega_\phi=0$ in the period of (scalar) dark matter domination (from $N\cong -7.5$ to $N\cong -1$ where $N=\ln a$ is the e-folding number) as in the standard $\Lambda$CDM model.

We observe that  $\omega_{\phi}\rightarrow-1/3$ from above in the past, thus there are no further acceleration periods due to the scalar field in the past.  In addition, for $\alpha<0$, $\omega_{\phi}$ decreases after the matter domination period, while for $\alpha=0$ or $\alpha=1$, $\omega_{\phi}$ increases again. We also compute the age of the Universe in the different cases, compared to the $\Lambda$CDM case. The results are reported in the Table \ref{tb:CH(aCS)_age_universe}.
\begin{table}[htpb]
\begin{tabular}{ l | c | c | c | r }
Model & $\alpha$ & $\beta$ & Age of the universe [Gyr] & Source\\
  \hline			\hline
(\ref{eq:CH(aCS)_action}) & $1$ & $1$ & $10.603$ &\code{hi\_class}\\
(\ref{eq:CH(aCS)_action}) & $1$ & $0.4$ & $9.710$ &\code{hi\_class}\\
(\ref{eq:CH(aCS)_action}) & $0$ & $1$ & $11.807$ &\code{hi\_class}\\
(\ref{eq:CH(aCS)_action}) & $-1$ & $1$ & $15.147$ &\code{hi\_class}\\
\hline
$\Lambda$CDM & $0$ & $0$ & $13.799 \pm 0.021$ & Planck 2015 experiment \cite{Planck_2015}\\  
\hline
\end{tabular}
\centering
\caption{Table of the age of the universe computed with \code{hi\_class} for different values of the action parameters $\alpha$ and $\beta$.}\label{tb:CH(aCS)_age_universe}
\end{table}

\section{Stability issues}
\label{sec:stability}

\begin{figure}[htpb]
  \vspace*{-0.7cm}\begin{center}
    \includegraphics[width=0.83\textwidth]{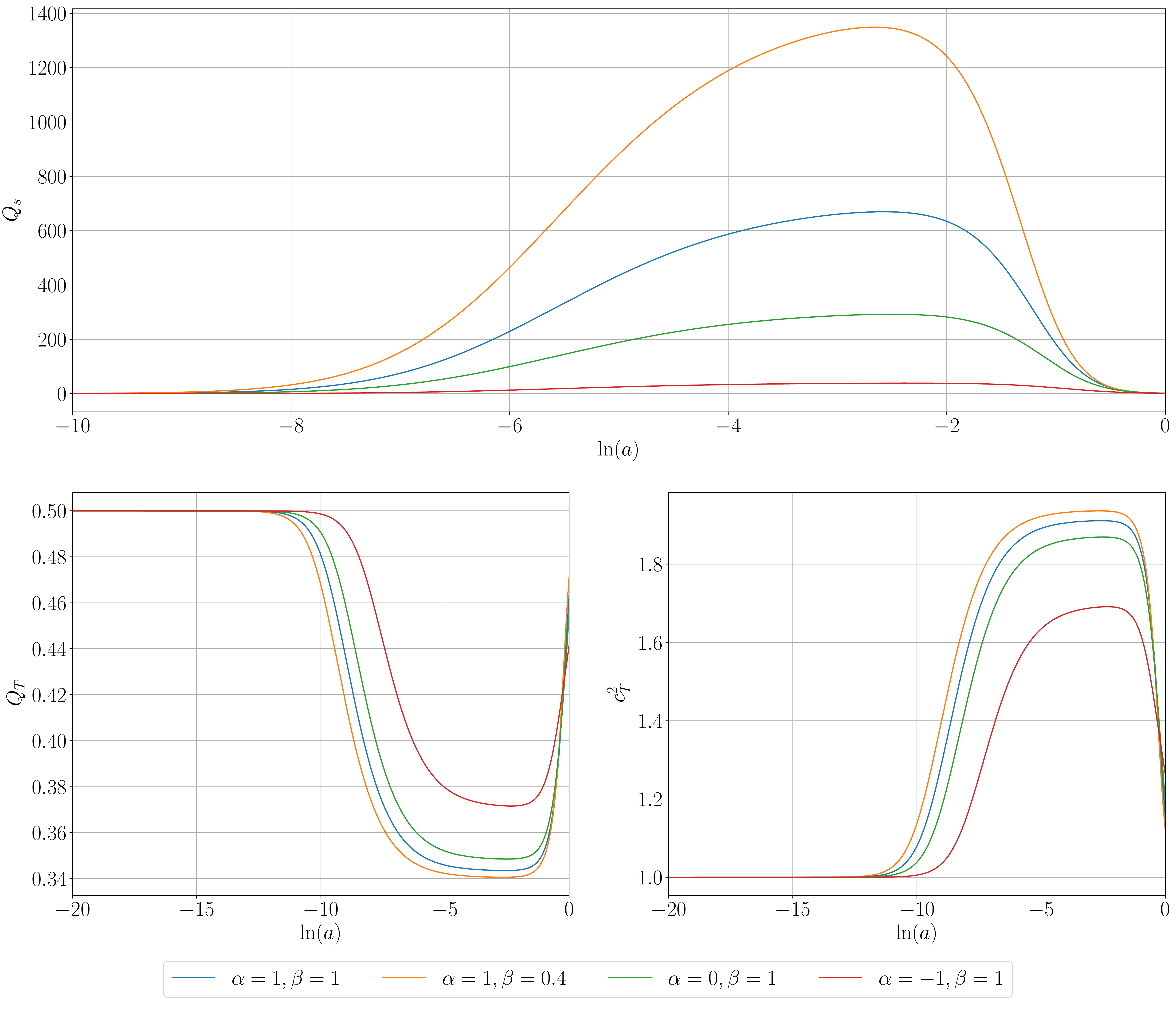}
  \end{center}
  \vspace*{-0.8cm}\caption{Plot of the functions associated with the stability conditions (\ref{eq:CH(MGR)_horndeski_Q_s}), (\ref{eq:CH(MGR)_horndeski_Q_T}) and (\ref{eq:CH(MGR)_horndeski_grav_waves_speed^2}). All these functions are positive during the whole evolution, and therefore they satisfy the stability conditions.}
  \label{fig:CH(aCS)_stability}
  \begin{center}
    \includegraphics[width=0.83\textwidth]{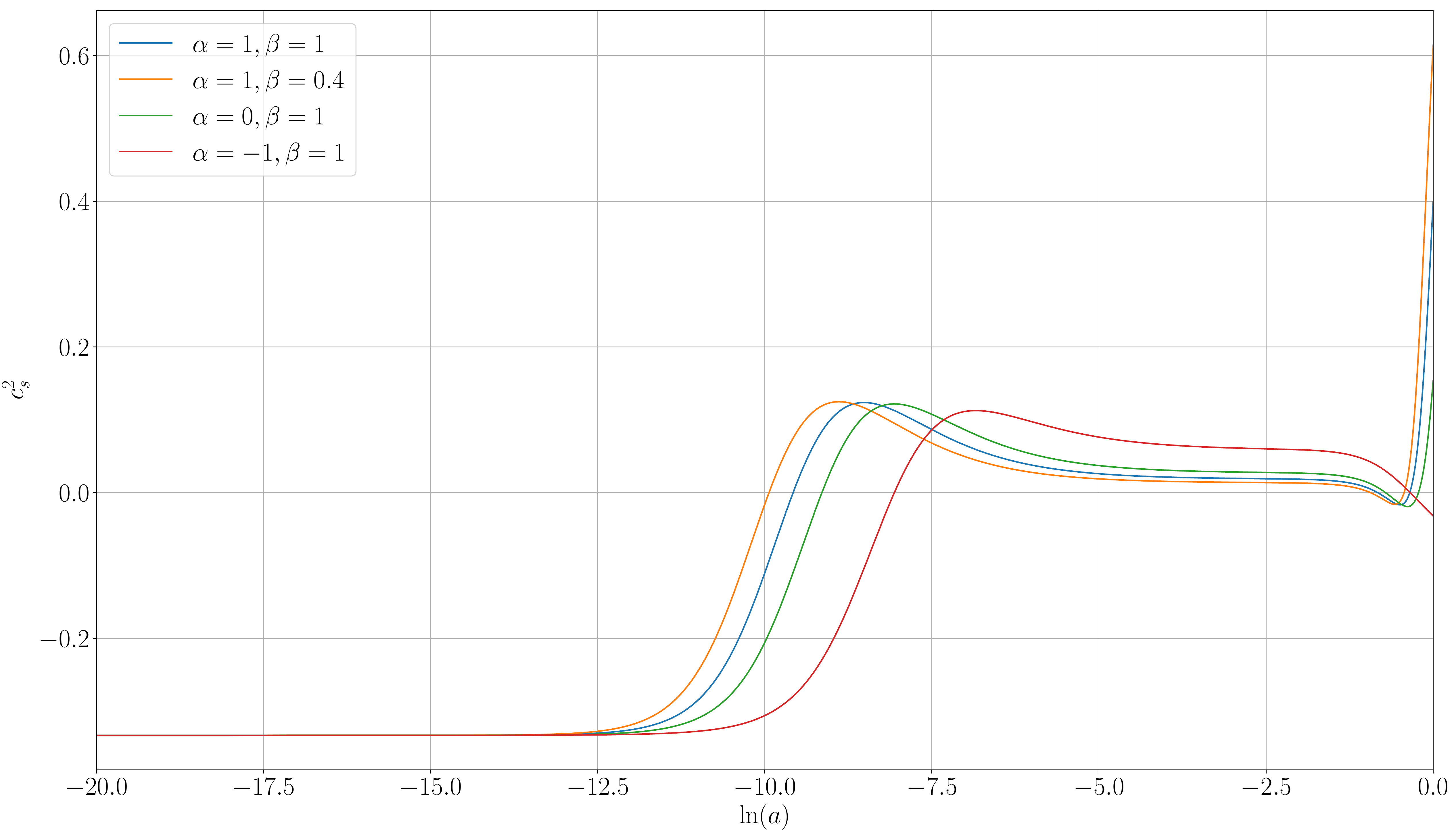}
  \end{center}
  \vspace*{-0.6cm}\caption{Plot of the scalar sound speed squared (\ref{eq:CH(MGR)_horndeski_sound_speed^2}). The asymptotic value at early times is a negative value, i.e. $c_s^2=-1/3$.}
  \label{fig:CH(aCS)_sound_speed}
\end{figure}

\noindent Before turning to the numerical  perturbative analysis, it is useful to check the stability of the background solutions. To do so,  we use the formalism of \cite{Zumalacarregui:2016pph} (see also \cite{tsuji}) and we consider the conditions
\begin{align}
Q_s &\equiv \frac{2 M_*^2 D}{(2-\alpha_B)^2}>0\label{eq:CH(MGR)_horndeski_Q_s}\\
c_s^2 &\equiv \frac{1}{D} \left[\left(2-\alpha_B\right)\left(-\frac{\dot{H}}{aH^2}+\frac{1}{2}\alpha_B \left(1+\alpha_T\right)+\alpha_M-\alpha_T\right)-\frac{3\left(\rho+P\right)}{H^2 M_*^2}+\frac{\dot{\alpha_B}}{aH}\right]>0\label{eq:CH(MGR)_horndeski_sound_speed^2}\\
Q_T &\equiv \frac{M_*^2}{8}>0\label{eq:CH(MGR)_horndeski_Q_T}\\
c_T^2 &\equiv 1+\alpha_T>0,\label{eq:CH(MGR)_horndeski_grav_waves_speed^2}
\end{align}
where $D\equiv \alpha_K+ 3\alpha_B^2/2$, and $\rho$ and $P$ are respectively the total energy density and the pressure of the matter content, excluding the contribution from the scalar field. The conditions (\ref{eq:CH(MGR)_horndeski_Q_s}) and (\ref{eq:CH(MGR)_horndeski_sound_speed^2}) are related to the scalar perturbations (subscript $s$), while (\ref{eq:CH(MGR)_horndeski_Q_s}) and (\ref{eq:CH(MGR)_horndeski_grav_waves_speed^2}) are related to the tensor perturbations (subscript $t$). In order to avoid instabilities, we require that these conditions are satisfied at all times. The functions $Q_s$, $Q_T$ and $c_T^2$ are plotted in Fig.\  \ref{fig:CH(aCS)_stability}: we see that they are always positive and thus they satisfy the stability conditions.

On the contrary, a potential problem arises in the scalar sound speed squared (\ref{eq:CH(MGR)_horndeski_sound_speed^2}), that might lead to instabilities in the perturbations. In fact, from Fig.\ \ref{fig:CH(aCS)_sound_speed}, we see that $c_{s}^{2}$ becomes negative in two epochs.

Let us first analyse the more problematic instability at early times.
By expanding $c_{s}$ near $a=0$ (or $N\rightarrow - \infty$) we find
\begin{align}
c_s^2 &= -\frac{1}{3}+\frac{4 \Omega_\phi  (3 \beta +\alpha  \Omega_\Lambda^0)^2}{9 \beta  (\Omega_r^0)^2 (9 \beta +\alpha  \Omega_\Lambda^0)} a^2  -\frac{5 \alpha  \Omega_\Lambda^0}{6 \beta  \Omega_r^0} a^4 + {\cal O}\left(a^5\right) \label{eq:CH(aCS)_cs2_expansion}\\&= -\frac{1}{3}+\frac{2 \Omega_\Lambda^0 q^2}{27 H_0^2 (\Omega_r^0)^2 \beta} a^2 -\frac{5 \alpha \Omega_\Lambda^0}{6 \beta  \Omega_r^0}a^4 + {\cal O}\left(a^5\right) \nonumber,
\end{align}
where we have considered only radiation as matter content (here, the superscript $0$ indicates quantities calculated at the present time), so $H=H_0 \sqrt{\Omega_r}/a^2$. The second equality is computed substituting $\Omega_\phi$ with (\ref{eq:CH(aCS)_fractional_density_analytical}). We readily see that $c_s^2$ is negative at $a=0$, and therefore it does not satisfy the above condition (\ref{eq:CH(MGR)_horndeski_sound_speed^2}) at all times.

In principle, one can push back the times at which $c_{s}^{2}$ becomes negative until the model is no longer valid as the Universe is in the inflationary phase.  From (\ref{eq:CH(aCS)_cs2_expansion}) it is obvious that, in order to make the sound speed positive at the beginning of  our post-inflationary computation (numerically, we set $a_{\text{start}}=10^{-14}$) we need the condition $a_{\text{start}}^2 q^2/H_0^2\beta \gtrsim 1$, that is
\begin{equation}
\beta \lesssim q^2 10^{-20}\,.
\label{eq:CH(aCS)_beta_cs2_condition}
\end{equation}
The integration constant $q^2$ is usually small because it is directly related to the value of the scalar field at $a_\text{start}$, which cannot be too large (otherwise $\phi$ becomes dominant at $a_\text{start}$). Therefore we would need a very small $\beta$ in order to have a positive $c_s^2$ and cancel the instability. But $\beta$ small means that the action (\ref{eq:CH(aCS)_action}) becomes a Quintessence action with a null potential. This is potentially a problem, because the state parameter in a Quintessence model with null potential leads to a $\omega_\phi$ equal to $1$ from the beginning of the evolution. This means that the scalar field would be either dominant or, if $q$ is small enough, the scalar field would be not dominant but still have an non-physical value of $\omega_\phi$ if we want to mimic dark matter. 

Note that the violation of condition (\ref{eq:CH(MGR)_horndeski_sound_speed^2}) does not automatically means that the model will be affected by instabilities, since the perturbation differential equations stability conditions consider also other terms, as we will see in equation (\ref{eq:CH(aCS)_perturbation_differential_equation}). We will further investigate the implications of these results in the next sections. Finally, there is a second instability at recent times (near $N\cong 1$). As we will see in section \ref{sec:CH(aCS)_stability_disc}, this instability can also cause instabilities in the perturbations.

\section{Numerical analysis of the perturbations}\label{sec:perturbations}

\noindent We now turn to the main results of this paper, namely the ones concerning linear perturbations calculated with the code \code{hi\_class}. In particular, we analyse the CMB and the matter power spectrum. In the following, we will consider not only the case when the scalar field simulates the cosmological dark matter fluid but also the case when it is sub-dominant at the background level (i.e. dark matter is not the scalar field).

\subsection{CMB and matter power spectrum}
\label{sec:CH(aCS)_CMB_MPSM}

\noindent Let us consider the following situations:
\begin{enumerate}[label=(\roman*)]
\item \label{list:CH(aCS)_enumerate_case1} the same $\alpha$ and $\beta$ of the previous section, i.e.\ the scalar field accounts for dark matter entirely. You can immediately see in Fig.\ \ref{fig:CH(aCS)_Pk_mimicking_darkmatter} the instability in the matter power spectrum of the matter at almost every scale of interest (the values of $k$ considered in Fig. \ref{fig:CH(aCS)_Pk_mimicking_darkmatter}). 

\begin{figure}[H]
  \begin{center}
    \includegraphics[width=1.0\textwidth]{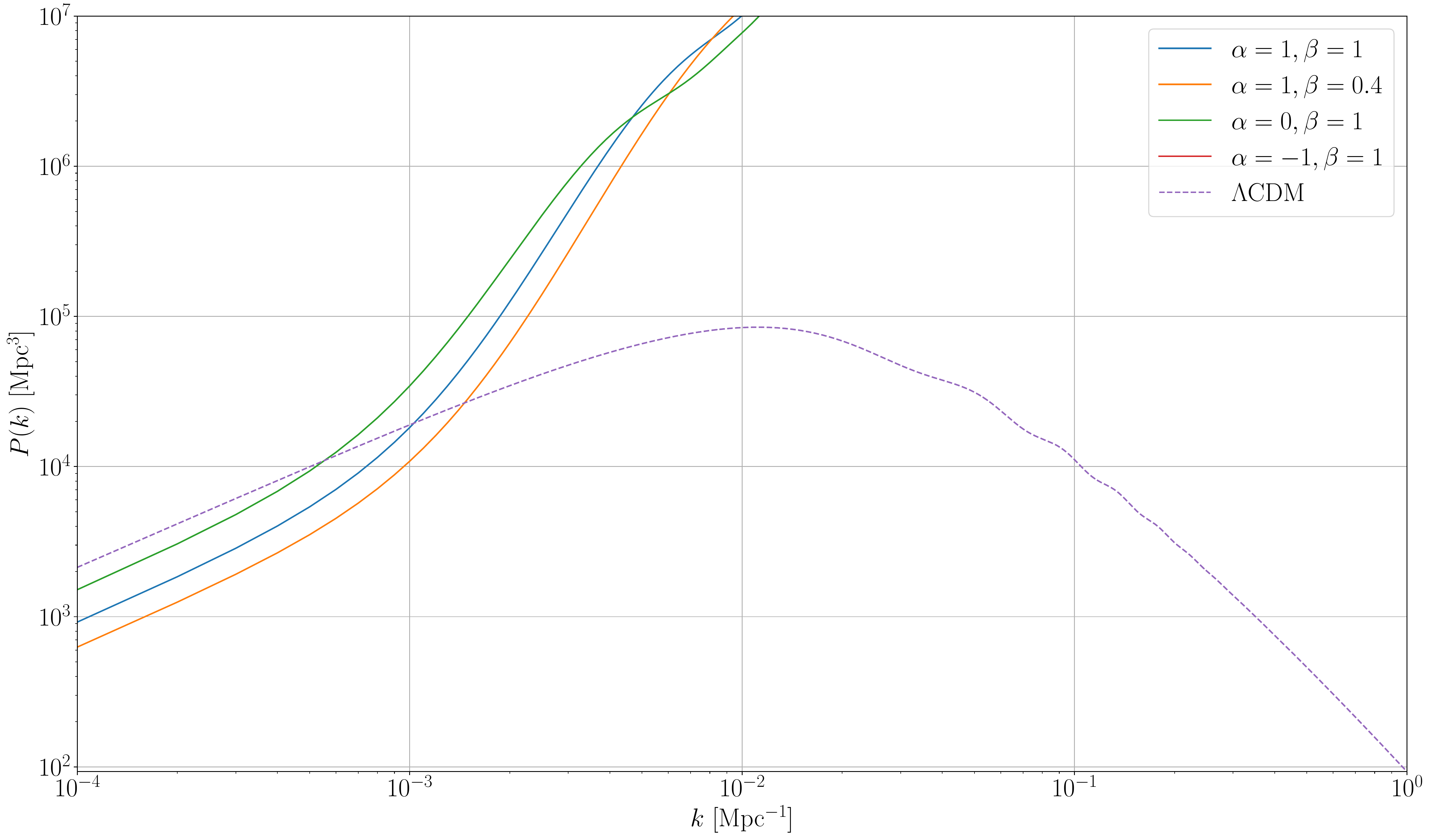}
  \end{center}
  \caption{Plot of the matter power spectrum in the case \ref{list:CH(aCS)_enumerate_case1}. The parameters are the same considered in section \ref{sec:aCS_background} that mimic the dark matter (see Fig.\ s \ref{fig:CH(aCS)_rinaldi_background} and \ref{fig:CH(aCS)_rinaldi_background2}).}
  \label{fig:CH(aCS)_Pk_mimicking_darkmatter}
\end{figure}
\item \label{list:CH(aCS)_enumerate_case2} the case with $\alpha=1$ and a very small $\beta=10^{-10}$, $\Omega_{cdm} h^2 = 0.1197$ (as in $\Lambda$CDM). In this case, the effects of the scalar field should be negligible at the perturbation level. As previously mentioned in the analysis of $c_s^2$, if we want to use a small $\beta$, we also need a small $\psi \cong 10^{-10}$, otherwise the scalar field will dominate the other components from the beginning. Thus, the derivative of the scalar field is very small also at the background level which is basically the same as $\Lambda$CDM. Nevertheless, although the scalar field is sub-dominant, it is enough to cause divergences in the perturbations, as one can see in Fig. \ref{fig:CH(aCS)_Pk_small_change} and, to a lesser extent, in Fig.\  \ref{fig:CH(aCS)_CMB_small_change}. The instabilities are at small scales (large $k$), and therefore they might be produced by a negative sound speed (gradient instabilities), as explained below.

\begin{figure}[H]
  \begin{center}
    \includegraphics[width=1.0\textwidth]{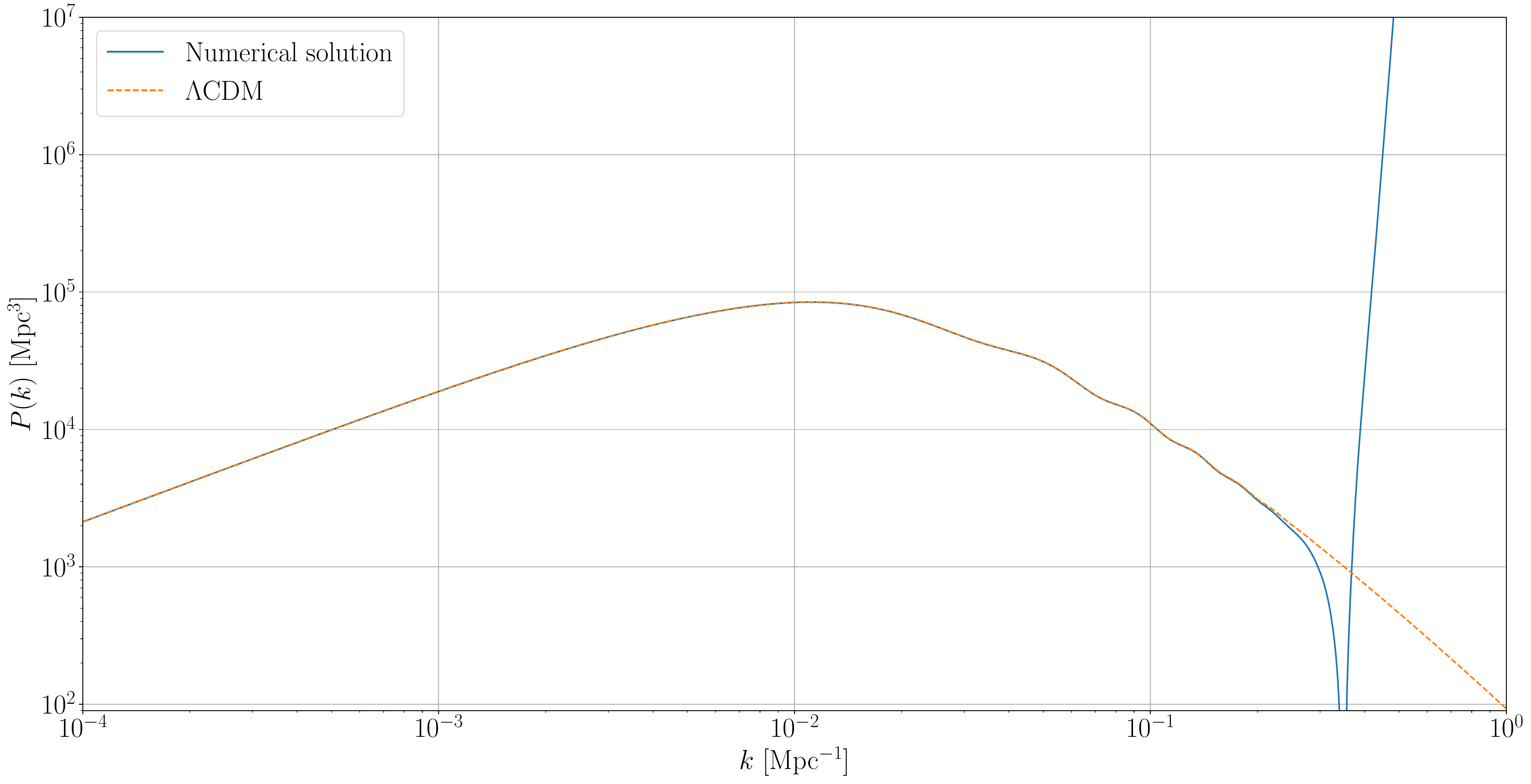}
  \end{center}
  \caption{Plot of the matter power spectrum in the case \ref{list:CH(aCS)_enumerate_case2} where $\alpha=1$, $\beta=10^{-10}$, $\Omega_{cdm} h^2 = 0.1197$ and $\phi$ modelling dark energy. In this plot the divergence at small scales (large $k$) is clearly evident.}
  \label{fig:CH(aCS)_Pk_small_change}
\begin{center}
    \includegraphics[width=1.0\textwidth]{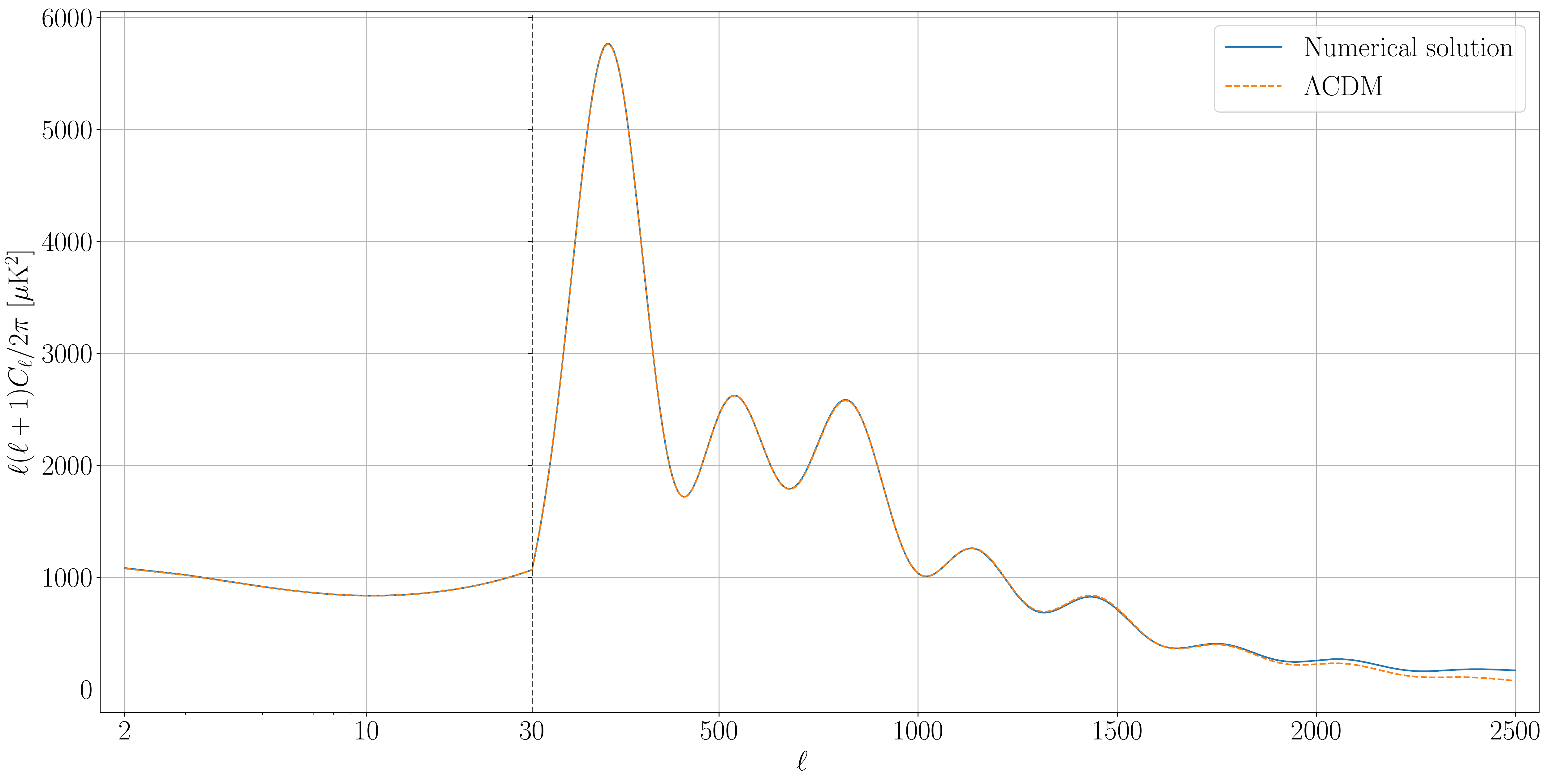}
  \end{center}
  \caption{Plot of the CMB in the case \ref{list:CH(aCS)_enumerate_case2} where $\alpha=1$, $\beta=10^{-10}$, $\Omega_{cdm} h^2 = 0.1197$ and $\phi$ is modelling dark energy. We plot the non-lensed $C_\ell$. In this plot the diverge is not evident, but you can see the slightly difference of the model with respect to $\Lambda$CDM at small scales (large $\ell$).}
  \label{fig:CH(aCS)_CMB_small_change}
\end{figure}
\begin{figure}[H]
  \begin{center}
    \includegraphics[width=1.0\textwidth]{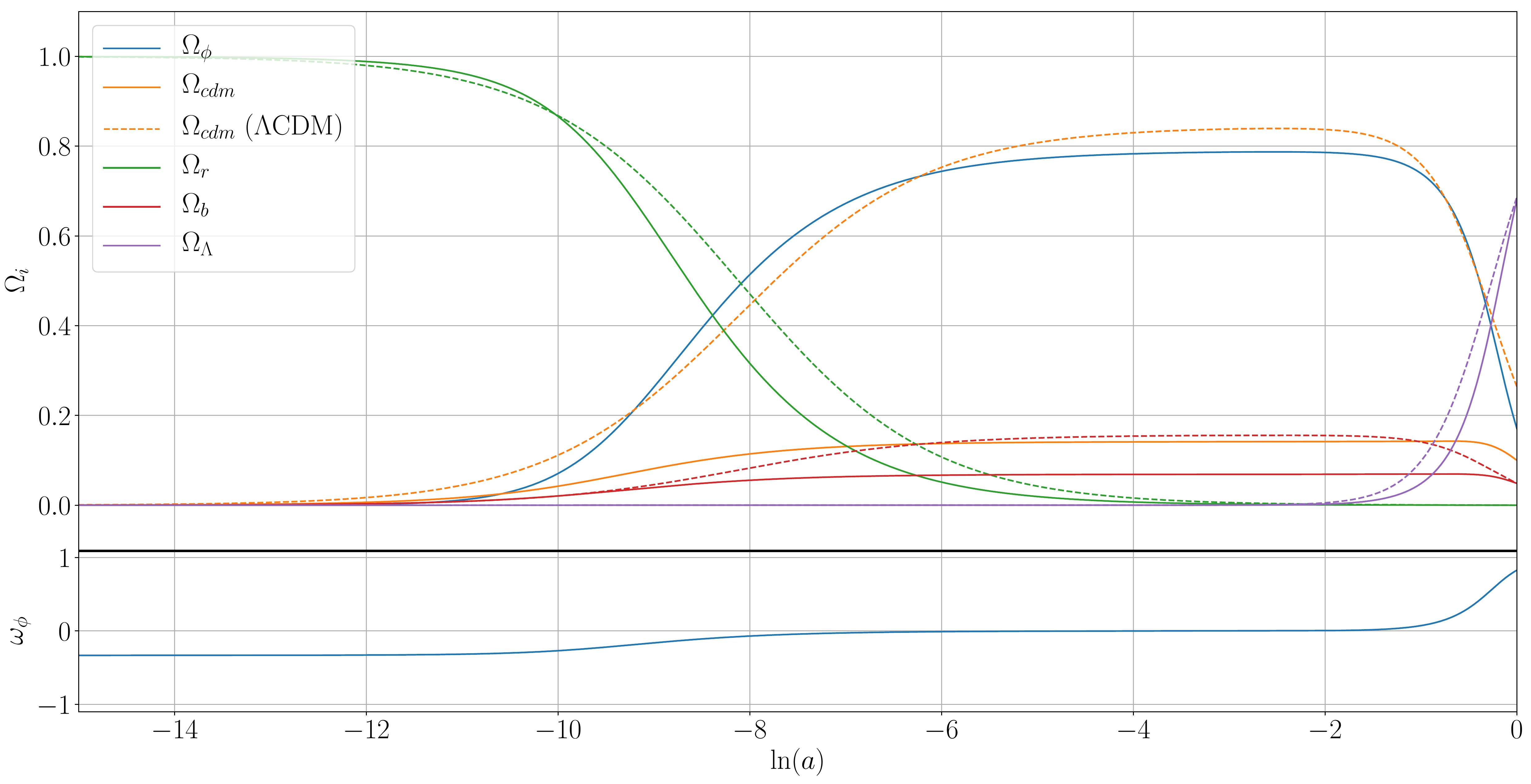}
  \end{center}
  \caption{Plot of the background in the case \ref{list:CH(aCS)_enumerate_case3} where $\Omega_{cdm}^0=0.1$. The dashed lines are the $\Lambda$CDM densities. Since the state parameter $\omega_\phi=0$, the scalar field behaves like the dark matter in the matter domination period, but has smaller fractional densities values with respect to the $\Omega_{cdm}^0=0$ case (Fig.\  \ref{fig:CH(aCS)_rinaldi_background}) due to the presence of a cold dark matter fraction $\Omega_{cdm}^0=0.1$.}
  \label{fig:CH(aCS)_bkg_part_cdm}
  \begin{center}
    \includegraphics[width=1.0\textwidth]{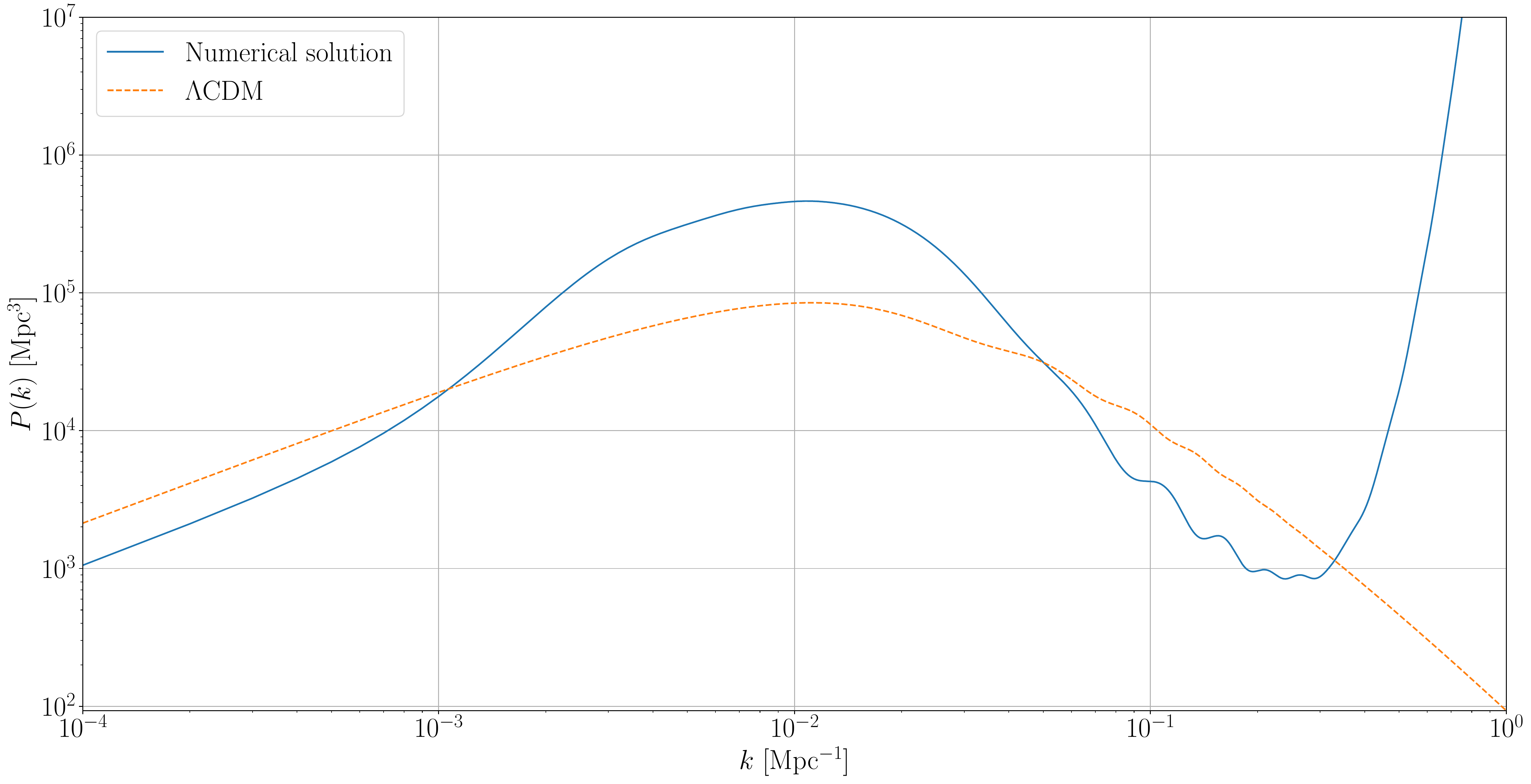}
  \end{center}
  \caption{Plot of matter power spectrum in the case \ref{list:CH(aCS)_enumerate_case3} where $\Omega_{cdm}^0=0.1$.}
  \label{fig:CH(aCS)_Pk_part_cdm}
\end{figure}

\item \label{list:CH(aCS)_enumerate_case3}  the case $\alpha=1$ and $\beta=1$, but with the addition some cold dark matter, i.e. $\Omega_{cdm}^0=0.1$ instead of $0$. Therefore, here we assume that the scalar field contributes only partially to the dark matter content of the Universe. The background evolution is shown in Fig.\  \ref{fig:CH(aCS)_bkg_part_cdm}. Also in this case divergences appear at small scales, see Fig.\  \ref{fig:CH(aCS)_Pk_part_cdm}.
\end{enumerate}

\subsection{Analysis of divergences}
\label{sec:CH(aCS)_stability_disc}

\begin{figure}[H]
  \begin{center}
    \includegraphics[width=1.0\textwidth]{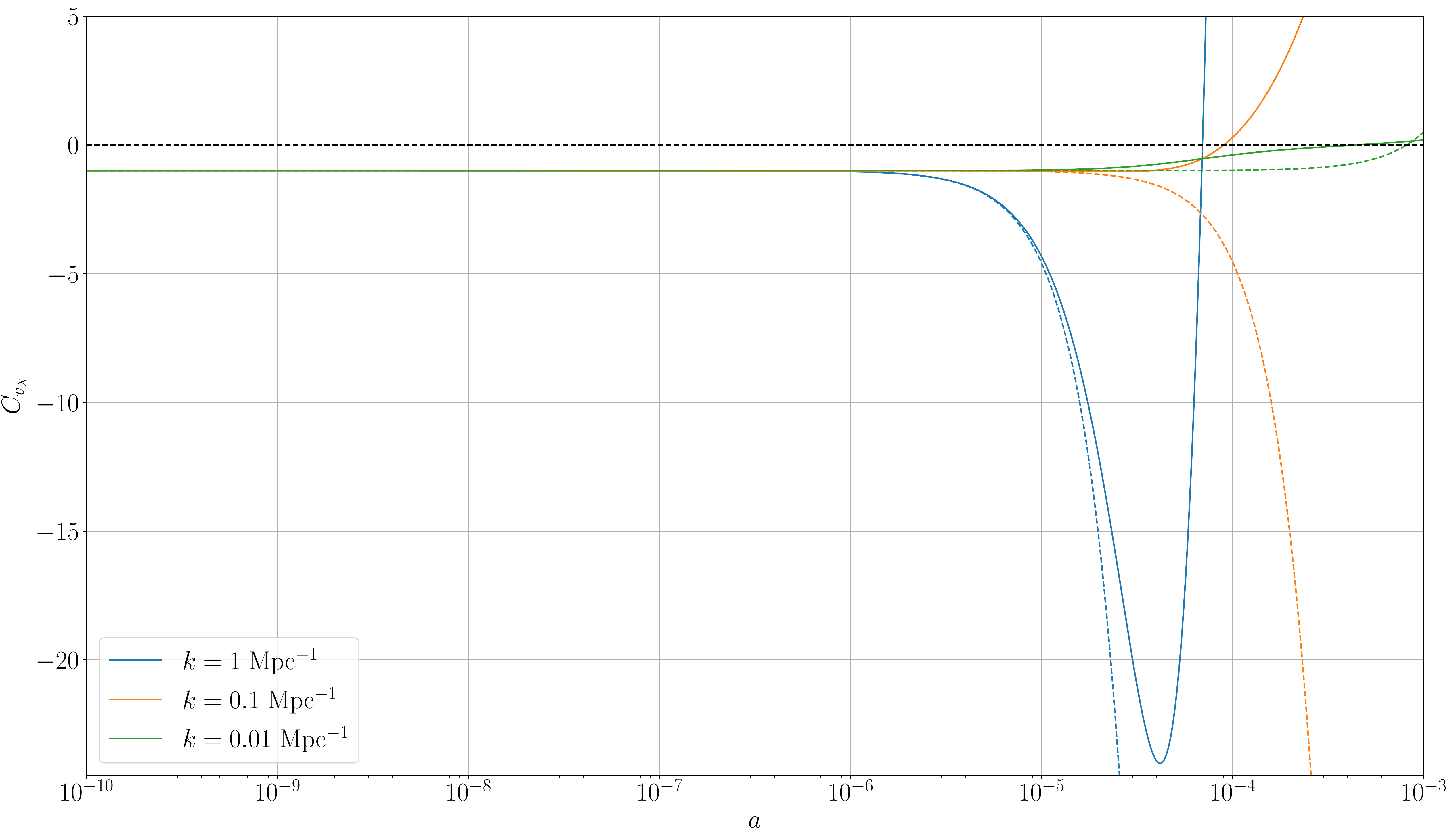}
  \end{center}
  \caption{Plot of $C_{v_X}$ in the $\alpha=1$ and $\beta=1$ case. The continuous lines are the numerical result for $C_{v_X}$, and the dashed lines are the series truncated at the second order in $a^2$. We can see that the truncated analytical series is good only for $a<10^{-5}$. Above the black dashed line, $C_{v_X}>0$ and the scalar field perturbations differential equation (\ref{eq:CH(aCS)_perturbation_differential_equation}) has an oscillatory behaviour.}
  \label{fig:CH(aCS)_difference_plot}
\end{figure}

\noindent In the previous section we have seen (Figs.\  \ref{fig:CH(aCS)_Pk_mimicking_darkmatter}, \ref{fig:CH(aCS)_Pk_small_change} and \ref{fig:CH(aCS)_Pk_part_cdm}) that the CMB and the matter power spectrum  are affected by a divergent behaviour at small scales. This problem can be partly explained by the fact that the sound speed $c_s^2$ is negative at early times. In order to explain this more thoroughly, we have to consider the differential equation for the scalar field perturbations (A.18) of \cite{Zumalacarregui:2016pph}. In synchronous gauge, it reads
\begin{equation}
\ddot{v}_X + A \dot{v}_X + \frac{2 a^2 H^2}{2-\alpha_B} \left(\frac{c_s^2 k^2}{a^2 H^2} - 4\frac{\lambda_8}{D}\right) v_X = F,
\label{eq:CH(aCS)_perturbation_differential_equation}
\end{equation}
where $c_s^2$ is defined in equation (\ref{eq:CH(MGR)_horndeski_sound_speed^2}), $k$ is the perturbation mode wavenumber, $D\equiv \alpha_K+ 3\alpha_B^2/2$, and  $v_X = \delta \phi / \phi$. The functions $A$ and $F$ depend on the $\alpha$ functions, but the exact form is irrelevant for our purposes. The function $\lambda_8$ is  given by (A.27) of \cite{Zumalacarregui:2016pph}, namely
\begin{equation}
\begin{aligned}
\lambda_8 = &- \frac{\lambda_2}{8} \left( D - 3\lambda_2 + \frac{3 \dot{\alpha}_B}{aH}\right)+\frac{1}{8}(2-\alpha_B)\left[(3\lambda_2-D)\frac{\dot{H}}{aH^2}-\frac{9\alpha_B\dot{P}_m}{2 a H^3 M_*^2}\right]\\
&-\frac{D}{8}(2-\alpha_B)\left[4 + \alpha_M + \frac{2\dot{H}}{a H^2} + \frac{\dot{D}}{a H D}\right].
\end{aligned}
\end{equation}
From equation (\ref{eq:CH(aCS)_perturbation_differential_equation}) we immediately see that in order to have a non-exponential behaviour, the condition $c_s^2>0$ is not sufficient. Rather,  we must impose the condition $C_{v_X}>0$, where
\begin{align}
C_{v_X}=\frac{1}{2-\alpha_B} \left(\frac{c_s^2 k^2}{a^2 H^2}- 4\frac{\lambda_8}{D}\right)\,.
\end{align}
As we did for the sound speed, we expand $C_{v_X}$ near $a=0$. To simplify, we assume also here that the Universe is dominated by radiation so $H=H_0 \sqrt{\Omega_r}/a^2$. We find 
\begin{align}\label{expans}
C_{v_X}&=-1 + \frac{ \left(\frac{15 \Omega_\phi  (3 \beta +\alpha  \Omega_\Lambda^0)^2}{\beta  (9 \beta +\alpha  \Omega_\Lambda^0)}-\frac{2 \Omega_r^0 k^2}{H_0^2}\right)}{12 \left(\Omega_r^0\right)^2}a^2 +{\cal O}\left(a^4\right)\\&=-1+\frac{5 \Omega_\Lambda^0 q^2-4 \beta  \Omega_r^0 k^2}{24 \beta  H_0^2 (\Omega_r^0)^2}a^2 +{\cal O}\left(a^4\right)\nonumber.
\end{align}
We see that, for small $a$, $C_{v_X}$ is negative and this leads to an exponential growth of  $v_X$. To confirm the analysis we plot $C_{v_X}$  in Fig.\  \ref{fig:CH(aCS)_difference_plot}. Note that the series truncated at the second order (dashed lines) in $a$ is not enough to approximate $C_{v_X}$ (continuous lines) after $a\cong10^{-5}$.

\begin{figure}[H]
  \begin{center}
    \includegraphics[width=1.0\textwidth]{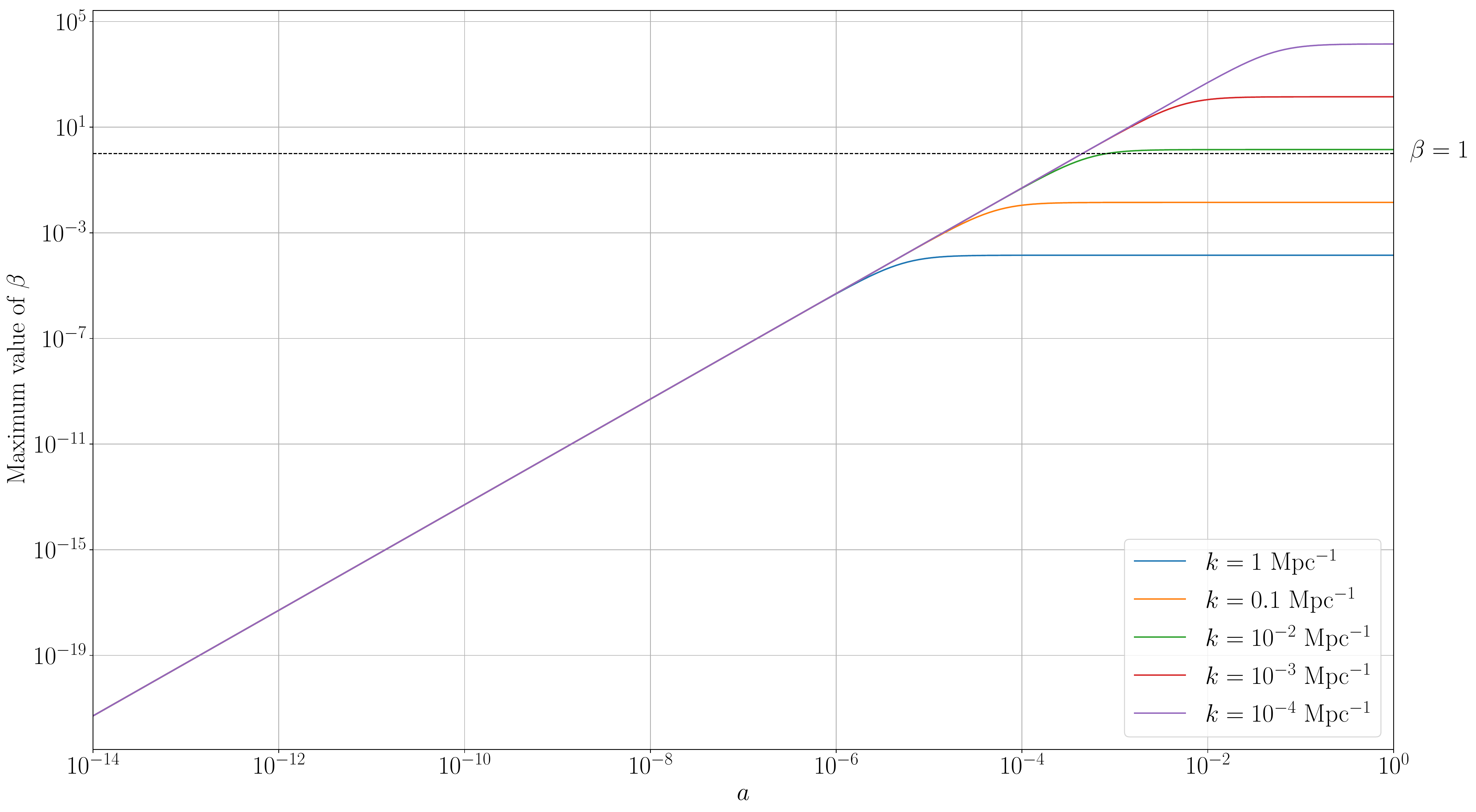}
  \end{center}
  \caption{Plot of the maximum value of $\beta$ allowed to avoid large deviations in the CMB and in the matter power spectrum, with $\alpha=1$ and $\beta=1$.}
  \label{fig:CH(aCS)_condition_beta}
\end{figure}

\noindent We can try to fix the instability choosing $q$ and $\beta$ such that the term proportional to $a^2$ in the expansion \eqref{expans} is equal or larger than $1$, i.e.
\begin{equation}
\frac{5 \Omega_\Lambda^0 q^2-4 \beta  \Omega_r^0 k^2}{24 \beta  H_0^2 (\Omega_r^0)^2}a^2 \geq 1.
\end{equation}
We can translate this  into the condition on $\beta$ given by 
\begin{equation}
\beta \lesssim \frac{5 q^2 \Omega_\Lambda^0}{24 a^{-2} H_0^2(\Omega_r^0)^2+4 k^2 \Omega_r^0} \,.
\label{eq:CH(aCS)_beta_condition_diff_eq_v_x}
\end{equation}
With this condition we can go back to the matter power spectrum computed in section \ref{sec:CH(aCS)_CMB_MPSM}. In particular, consider the case \ref{list:CH(aCS)_enumerate_case1} where the scalar field mimics the dark matter. If we consider the initial value of the scalar field found by \code{hi\_class} enforcing the sum rule (\ref{eq:CH(aCS)_sum_rule}), that is given by $q\cong10^{-4}$, we can compute the maximum value of $\beta$ allowed in order to avoid instabilities from the condition (\ref{eq:CH(aCS)_beta_condition_diff_eq_v_x}). Since the condition explicitly depends on the wavenumbers and on time (trough the scale factor $a$), in Fig.\  \ref{fig:CH(aCS)_condition_beta} we show the results for different $k$, as a function of the scale factor $a$. In the following paragraphs we will also explore the applicability limits of these results.

\noindent To better understand Fig. \ref{fig:CH(aCS)_condition_beta}, we consider the approximations regime plot of \code{CLASS} \cite{CLASS}, for which we present a version in the Appendix, in Fig.\  \ref{fig:CH(soft)_approx_class}. From this plot we see that the time at which the computation of the perturbations starts depends on the wavenumber $k$. For large scales (typically  $k\lesssim 10^{-2}$ Mpc$^{-1}$), the computation starts at a conformal time $\eta\cong 5 \times 10^2$ Mpc, that is $a\cong10^{-5}$. For smaller scales, the computation starts earlier with the earlier time being $\eta\cong10^2$ Mpc, that is $a\cong10^{-7}$. 

Therefore, for large scales, the condition (\ref{eq:CH(aCS)_beta_condition_diff_eq_v_x}) is not met for a short interval of time. Later on, these scales will inevitably fall in the regime where the condition is satisfied (above the $\beta=1$ horizontal line in the Fig.\  \ref{fig:CH(soft)_approx_class}, that is the value of $\beta$ chosen in order to mimic dark matter with the scalar field). On the contrary, small scales never satisfy the condition and therefore show an exponential growth during the epoch where the scalar field dominates, causing the large deviations that we see in the matter power spectrum in Fig.\  \ref{fig:CH(aCS)_Pk_mimicking_darkmatter}. A further confirmation of that is the fact that matter power spectrum diverges when $k > 10^{-2}$ Mpc$^{-1}$.

Note that the condition (\ref{eq:CH(aCS)_beta_condition_diff_eq_v_x}), and therefore our previous considerations, is only an approximation of the real condition after equality. In fact, we computed the condition on $\beta$ assuming $H$ dominated only by radiation, and this is obviously not true at late times. Moreover the value of $a$ is not small at late times and therefore higher orders in the series might become non negligible. This can also be seen in Fig.\  \ref{fig:CH(aCS)_difference_plot}, where we plotted the truncated analytical series at second order in $a$ from which we have computed the condition on $\beta$: after $a=10^{-5}$ the truncated analytical series (dashed lines) is not a good approximation of $C_{v_X}$ (continuous lines).

\begin{figure}[H]
  \begin{center}
    \includegraphics[width=1.0\textwidth]{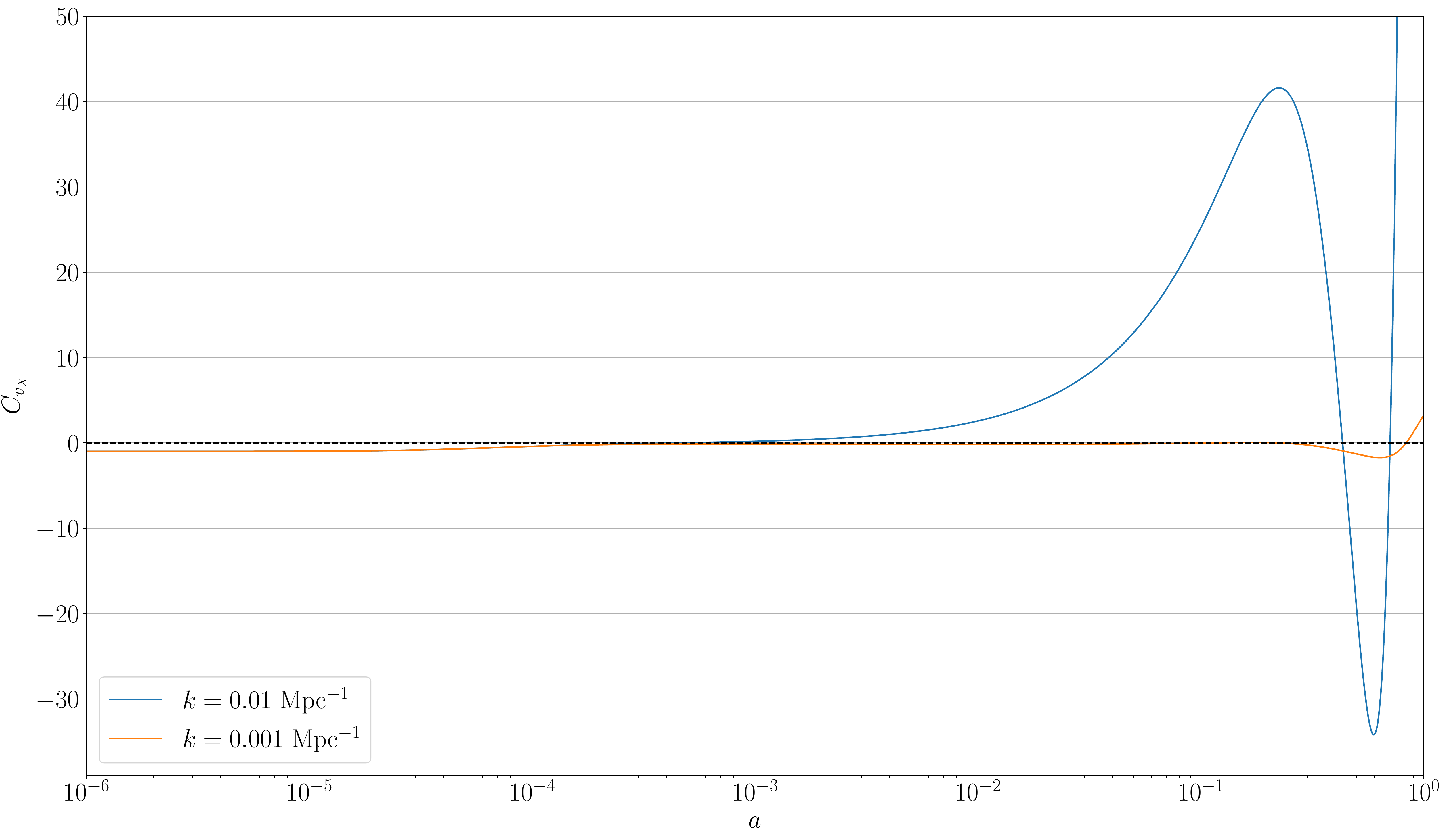}
  \end{center}
  \caption{Plot of $C_{v_X}$ in the $\alpha=1$ and $\beta=1$ case, for two modes. Above the black dashed line, $C_{v_X}>0$ and the scalar field perturbations differential equation (\ref{eq:CH(aCS)_perturbation_differential_equation}) has an oscillatory behaviour.  We see that the two modes have a second period where $C_{v_X}<0$ near today. Since this instabilities are well inside the period where \code{hi\_class} evaluates the perturbations, they can be source of instabilities in the perturbation functions.}
  \label{fig:CH(aCS)_difference_plot2}
\end{figure}

Finally, if we plot the entire numerical evolution of $C_{v_X}$ up to today, where our condition for $\beta$ is certainly not valid any more, we see that modes fall again in the exponential regime (where $C_{v_X}<0$). If the period where $C_{v_X}<0$ is long enough, this might introduces more instabilities even at large scales (small $k$), and therefore might produce divergences in the perturbation functions at all scales. The period where $C_{v_X}<0$ highly depends on the parameter $\beta$ chosen. This is explicitly shown in Fig.\  \ref{fig:CH(aCS)_difference_plot2} for two wave-numbers.

A similar analysis can be carried on in the other two cases \ref{list:CH(aCS)_enumerate_case2} and \ref{list:CH(aCS)_enumerate_case3}. Similarly to the previous case, we show the plots of $C_{v_X}$ for some modes in the Appendix, in Figs.\  \ref{fig:CH(aCS)_difference_plot_case2} and \ref{fig:CH(aCS)_difference_plot_case3}.

\section{Bounds on the gravitational waves velocity}\label{sec:gravity_waves}

\noindent In this section we want to analyse the model comparing the results with the recent experimental discoveries in the field. The LIGO/Virgo results have constrained the speed of gravity waves to be $c_T^2=1$ (where $c=1$), up to an error of approximately $10^{-15}$ \cite{Creminelli:2017sry}-\cite{Baker:2017hug}.

More precisely, in a generic Horndeski model the value of the gravity wave speed is given by
\begin{equation}
c_T^2=1+\alpha_T.
\end{equation}
Therefore the LIGO/Virgo experiment puts constraints on $\alpha_T$, which must satisfy
\begin{equation}
|\alpha_T| \lesssim 10^{-15}.
\label{eq:cT2_constraint}
\end{equation}
We recall that, in our model,
\begin{equation}
\alpha_T =\frac{2 \beta  X}{2\kappa{\Lambda}-\beta  X}.
\end{equation}

\begin{figure}[H]
  \begin{center}
    \includegraphics[width=1.0\textwidth]{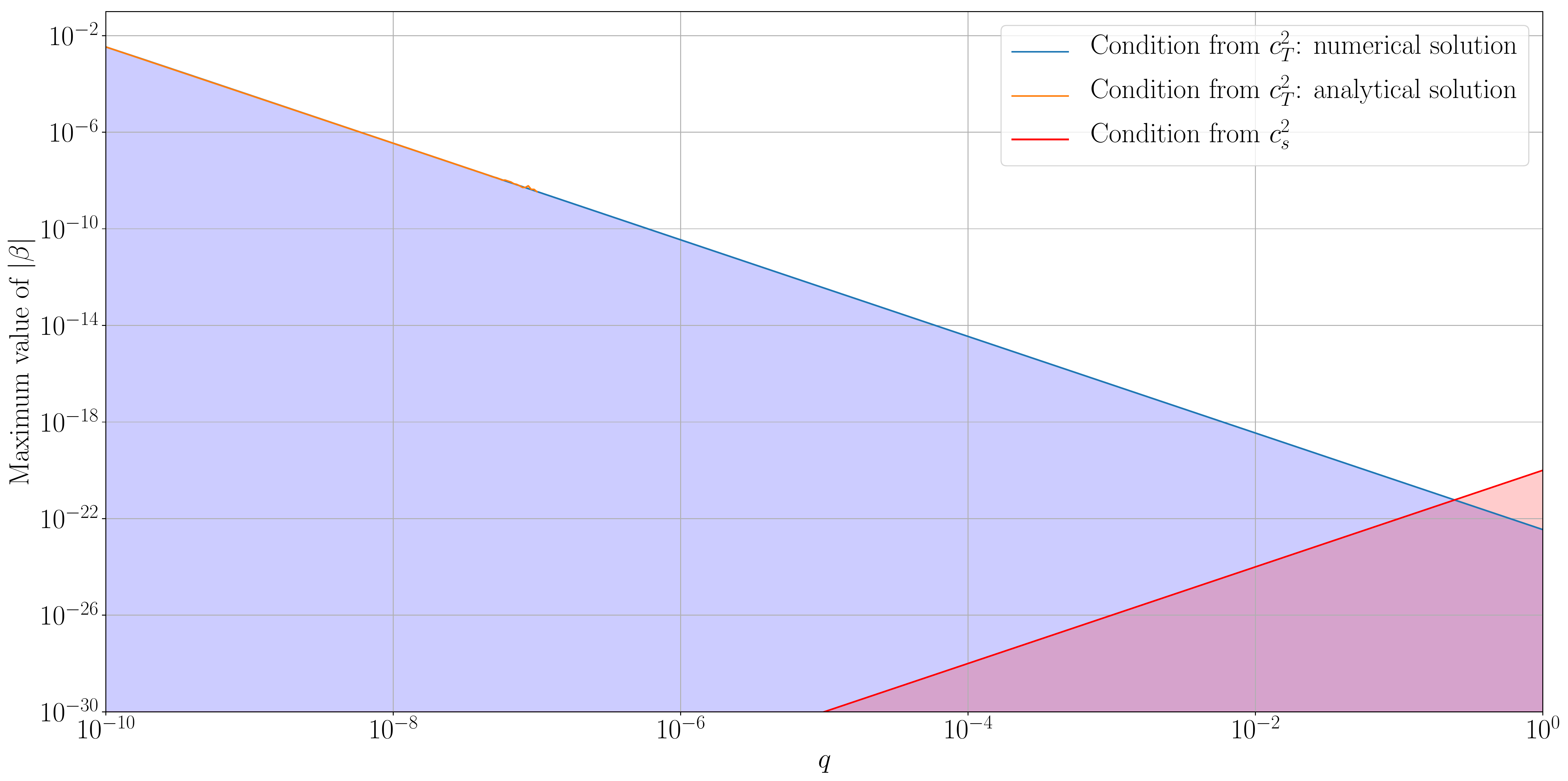}
  \end{center}
  \caption{Plot of the maximum value of $\beta$ allowed by the two constraints considered. The overlapping purple region is the one that must be considered in order to have a stable and non super-luminous gravity waves. The analytical solution for the condition on $c_T^2$ has been plotted only up to $\cong 10^{-7}$, since for higher $a$ it's affected by machine precision numerical instabilities.}
  \label{fig:CH(aCS)_plot_cond}
\end{figure}

For definiteness, we expand $\alpha_T$ near the late Universe ($a\sim 1$) in a model with $\beta\ll 1$ (since we know from the previous analysis on the perturbation differential equation that the case $\beta=1$ leads to instabilities). In this case the new scalar field does not mimic dark matter, and therefore we must add dark matter as the standard $\Lambda$CDM one in order to obtain a physically viable background. We find
\begin{align}
\alpha_T &= \frac{2 \beta  \Omega_\Lambda^0 q^2}{2 H_0^2 \left[\Omega_\Lambda^0 (\alpha +3 \beta )+3 \beta  \Omega_m^0\right]^2-\beta  \Omega_\Lambda^0 q^2}\label{eq:et_exp_alphaT}\\
&\phantom{=}-\frac{24 \beta  H_0^2 (\Omega_\Lambda^0)^2 q^2 (\alpha +3 \beta ) (\Omega_\Lambda^0 (\alpha +3 \beta )+3 \beta  \Omega_m^0)}{\left[\beta \Omega_\Lambda^0 q^2-2 H_0^2 (\Omega_\Lambda^0 (\alpha +3 \beta )+3 \beta  \Omega_m^0)^2\right]^2} (a-1) +{\cal O}\left[(a-1)^2\right].\nonumber
\end{align}
A trivial solution corresponds to $\alpha_T=0$, that is $\beta=0$. This is not unexpected since for $\beta=0$ our model reduces to quintessence with no potential, for which $c_T^2=1$ \footnote{See definition of $\alpha_T$ in \cite{Zumalacarregui:2016pph}.}.

In order to constrain $\beta$ with the LIGO/Virgo result (\ref{eq:cT2_constraint}), we first set the zero order term of the late time expansion (\ref{eq:et_exp_alphaT}) to be
\begin{equation}
\left|\frac{2 \beta  \Omega_\Lambda^0 q^2}{2 H_0^2 \left[\Omega_\Lambda^0 (\alpha +3 \beta )+3 \beta  \Omega_m^0\right]^2-\beta  \Omega_\Lambda^0 q^2}\right| < 10^{-15} \sim T
\label{eq:zero_order_term_condition}
\end{equation}
and we neglect the contribution of higher order terms. Using the fact that $\Omega^0_\Lambda + \Omega^0_m \cong 1$, and considering the conditions on $\beta$ found in the previous sections, we find the range of acceptable $\beta$ to be
\begin{equation}
|\beta| < \frac{\Omega^0_\Lambda q^2-6 \alpha  H_0^2 \Omega^0_\Lambda T-q\Omega^0_\Lambda \sqrt{q^2-12 \alpha H_0^2 T}}{18 H_0^2 T}.
\label{eq:beta_from_zero_order_term_condition}
\end{equation}
Note that the case $\alpha=0$ leads to the condition $\beta=0$, and therefore will not be considered in the following analysis.

We have also solved equation (\ref{eq:zero_order_term_condition}) numerically in order to find the maximum value of $\beta$ allowed, and the results are shown in Fig.\  \ref{fig:CH(aCS)_plot_cond}. We also show the maximum value of $\beta$ allowed by the constraint on $c_T^2$ computed with the analytical result (\ref{eq:beta_from_zero_order_term_condition}), and the one coming from the velocity of scalar perturbations $c_s^2$ (\ref{eq:CH(aCS)_beta_cs2_condition}), for $\alpha=-1$. The regime of allowed $\beta$ from the combination of the two conditions on $c_s^2$ and $c_T^2$ is the overlapping (purple) region in the plot.

In Fig.\  \ref{fig:CH(aCS)_plot_cond} we are considering $\beta$ as a function of $q$, which is the integration constant appearing in the definition of the scalar field (\ref{eq:CH(aCS)_field_solution_analytical}). This constant is computed by the software in order to satisfy the background condition (\ref{eq:CH(aCS)_sum_rule}). Moreover, from the discussion at the end of section \ref{sec:stability} we know that small $\beta$ means that $q$ found by the software is smaller. Therefore, the contribution of the scalar field on the background evolution is negligible, and the presence of the field becomes unnecessary for explaining the nature of dark matter and dark energy.

Again, this analysis has been carried on in the regime where $a\cong 1$, where the LIGO/Virgo experiment has been performed. In principle, we might want to check the condition on $\beta$ at earlier times to have $c_T^2=1$ at every times. Note that we have also neglected higher order terms in $(a-1)$ in the expansion (\ref{eq:et_exp_alphaT}) since they give smaller contributions to $\alpha_T$ with respect to the leading term in the expansion near $a=1$.

\section{Conclusions}

\noindent The model with the action (\ref{eq:CH(aCS)_action}) has an interesting feature for a certain range of parameters where $\beta \cong 1$: the scalar field mimics the behaviour of dark matter in the matter domination epoch, at least at the background level. In fact, we saw in the fractional densities and state parameter plots in Figs.\ \ref{fig:CH(aCS)_rinaldi_background} and \ref{fig:CH(aCS)_rinaldi_background2} that the scalar field has an equation of state parameter of $\omega_\phi=0$, the same of a cold dark matter component, during the matter domination epoch. In principle, varying the parameters of the model in a certain range, we could find the values of the parameters that correspond to the same fractional densities behaviour of $\Lambda$CDM.

Unfortunately, despite the similarity between the scalar field of this model and the dark matter of $\Lambda$CDM, the model is plagued with some instabilities in the perturbation functions. We argued that a possible explanation for these problems is the fact that the harmonic oscillator like equation for the scalar field perturbations $v_X$ (\ref{eq:CH(aCS)_perturbation_differential_equation}) has a coefficient for the term proportional to $v_X$ that is negative at early times, and in a lesser extent at late times. Since the numerical evaluation of this differential equation starts earlier for small scales, these are more affected by instabilities. It is evident that the origin of the instabilities is related to a more complex mechanism that the violation of the condition $c_{s}^{2}>0$, as discussed in sec.\ \ref{sec:CH(aCS)_stability_disc}.

In principle, we can take a small value of $\beta$ to avoid instabilities and also to accommodate the results of the LIGO/VIRGO detection of the neutron star merger GW170817. However, this means that the term in the action that mimics dark matter becomes negligible, and therefore the physical interesting features of the model fade away. Moreover, we have seen that as we decrease $\beta$ we approach the limit where the model is a Quintessence model with no potential, that only predicts an unphysical fluid component with the equation of state parameter $\omega = 1$. 

Another way to solve the instability problem might be to add new terms to the action, at the price of changing the physical features of the model. In this case, the addition of new terms might change the properties we need to formulate a viable description of neutron stars and rotational curves of galaxies, and therefore a new investigation of this phenomena should be carried on after a change of the action. Also, the exact choice of the terms that can remove the instabilities is a non trivial problem.

We wish to stress that the GW170817 bound on the speed of gravitational waves is related to a single measurement, at a specific time $a$ and scale $k$. As pointed out in the recent works \cite{deRham:2018red,Battye:2018ssx}, it is possible that the speed of gravitational waves is  very close to unity only at the scales probed by the  LIGO/Virgo experiment and might change at  cosmological scales, which are the ones of our interest. However, the instabilities found in the perturbations function are independent of the speed of gravitational waves, as we can see from equation \eqref{eq:CH(aCS)_perturbation_differential_equation}, and are always present for the  parameter range that allows the theory to mimic dark matter. Therefore, our conclusion that the Horndeski-like action (\ref{eq:CH(aCS)_action}) is not acceptable as a cosmological model of dark matter, is unchanged.


\textbf{Acknowledgements}: We are grateful to Emilio Bellini for helping us with the implementation of the model in the \code{hi\_class} code.

\appendix
\section{Numerical evaluation with $\alpha$ functions} \label{sec:numerical_evaluation_alphas}

\begin{figure}[H]
\centering
\resizebox{17cm}{3cm}{\begin{tikzpicture}[snake=zigzag, line before snake = 0mm, line after snake = 2mm]
\node (begin) at (-8,0)[align=center, color=black] {$H_{\text{old}} = \sum_{i, \text{ no }\phi} \rho_i$};
\draw[->] (-8,0.5) to [bend left=40] (-4.1,0.5);
\node (first) at (-4,0)[align=center, color=black] {$\rho_{\phi,\text{ old}}=\rho_\phi (H_{\text{old}})$};
\draw[->] (-3.9,0.5) to [bend left=40] (-0.1,0.5);
\node (second) at (0,0)[align=center, color=black] {$H_\text{new} = \sum_i \rho_{i}$};
\draw[->] (0.1,0.5) to [bend left=40] (3.9,0.5);
\node (third) at (4,0)[align=center, color=black] {$\rho_{\phi,\text{ new}}=\rho_\phi (H_{\text{new}})$};
\draw[<-] (-3.9,-0.5) to [bend right=15] (2.0,-0.7);
\node (second) at (-1,-1.6)[align=center] {if \textbf{not} (\ref{eq:CH(aCS)_condition_self_consistent}), go back with $H_{\text{new}}$ as $H_{\text{old}}$};
\draw[->] (6,0) to (8,0);
\node (second) at (7,0.5)[align=center] {if (\ref{eq:CH(aCS)_condition_self_consistent})};
\node (third) at (9.6,0)[align=center, color=black] {$\rho_{\phi,\text{ new}}$\\ is the \textbf{output}};
\draw [decorate,decoration={brace,amplitude=10pt,mirror,raise=1pt},yshift=10pt] (-1.5,-0.55) -- (5.9,-0.55) node [black,midway,xshift=0.8cm]{};
\end{tikzpicture}}
\caption{Scheme of the self-consistent procedure for finding $\rho_\phi$. The scheme for $P_\phi$ is similar (also with $\dot{H}$).}
\label{fig:CH(aCS)_self_consistent_scheme}
\end{figure}
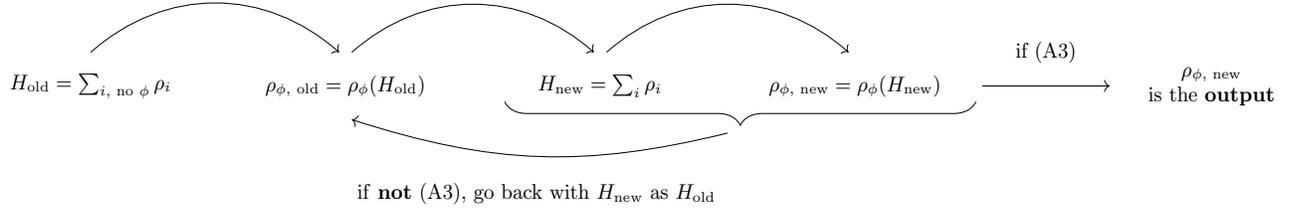

\noindent In this Appendix we discuss  the numerical algorithm used to compute the $\alpha$ functions. We have seen in section \ref{sec2} that we can write the $\alpha$ functions using background quantities only ($a$, $H$ and $\dot{H}$) because we have the analytic expression for $\phi$. But this is not enough to solve the parametrization problem. In fact, the $\alpha$ functions, $\rho_\phi$, and $P_\phi$ are still functions of $H$ and $\dot{H}$, that are computed by (\code{hi\_})\code{class} respectively with (in units of \code{hi\_class})
\begin{equation}
H^2 = \sum_{\text{species }i} \rho_i,
\label{eq:CH(soft)_Hubble_parameter}
\end{equation}
and
\begin{equation}
\dot{H} = - \frac{3 a}{2} \sum_{\text{species } i} \left(\rho_i + P_i\right).
\label{eq:CH(soft)_Hubble_parameter_prime}
\end{equation}
The equations (\ref{eq:CH(soft)_Hubble_parameter}) and (\ref{eq:CH(soft)_Hubble_parameter_prime}) are two sums over all densities and pressures that include the scalar field density $\rho_\phi$ and pressure $P_\phi$, which are functions of $H$ and $\dot{H}$ themselves. Therefore, if we use $H$ and $\dot{H}$ computed without the scalar field density to find $\rho_\phi$, we are introducing an error that can be non-negligible in the period where the scalar field is dominating the other components. A simple solution is to use a self-consistent procedure, as in the scheme in Fig.\  \ref{fig:CH(aCS)_self_consistent_scheme}, that is:
\begin{enumerate}[label=(\roman*),noitemsep]
\item compute $H_{\text{old}}=H$ and $\dot{H}=\dot{H}_{\text{old}}$ as a sum of the densities (and the pressures) of all the components except the scalar field density;
\item compute $\rho_{\phi, \text{ old}}$ and $P_{\phi, \text{ old}}$ with $H_{\text{old}}$ and $\dot{H}_{\text{old}}$. Then compute $H_{\text{new}}$ and $\dot{H}_{\text{new}}$, inserting $\rho_\phi$ and $P_\phi$ in the sums;\label{list:CH(aCS)_self-consistent_step2}
\item compute $\rho_{\phi, \text{ new}}$ and $P_{\phi, \text{ new}}$ with the new estimation of $H=H_{\text{new}}$ and $\dot{H}=\dot{H}_{\text{new}}$;
\item if the condition
\begin{equation}
\frac{\left|\rho_{\phi, \text{ new}}-\rho_{\phi, \text{ old}}\right|}{\rho_{\phi, \text{ old}}}<\epsilon
\label{eq:CH(aCS)_condition_self_consistent}
\end{equation}
where $\epsilon$ is an arbitrary small precision parameter, is not satisfied then go back to step \ref{list:CH(aCS)_self-consistent_step2} considering $\dot{H}_{\text{new}}$ as $H_{\text{old}}$ and $\dot{H}_{\text{new}}$ as $\dot{H}_{\text{old}}$. Otherwise stop the iterative process and consider as outputs $\rho_{\phi, \text{ new}}$, $P_{\phi, \text{ new}}$, $H_{\text{new}}$ and $\dot{H}_{\text{new}}$.
\end{enumerate}

With this solution, using $\epsilon = 10^{-6}$, we find that the $G_i$ functions method and the $\alpha$ functions method show a maximum relative difference of $0.1$\% in the fractional densities (see Appendix \ref{sec:equivalence_of_methods}).

\section{Numerical - Analytical comparison}

\begin{figure}[htpb]
  \begin{center}
    \includegraphics[width=1.0\textwidth]{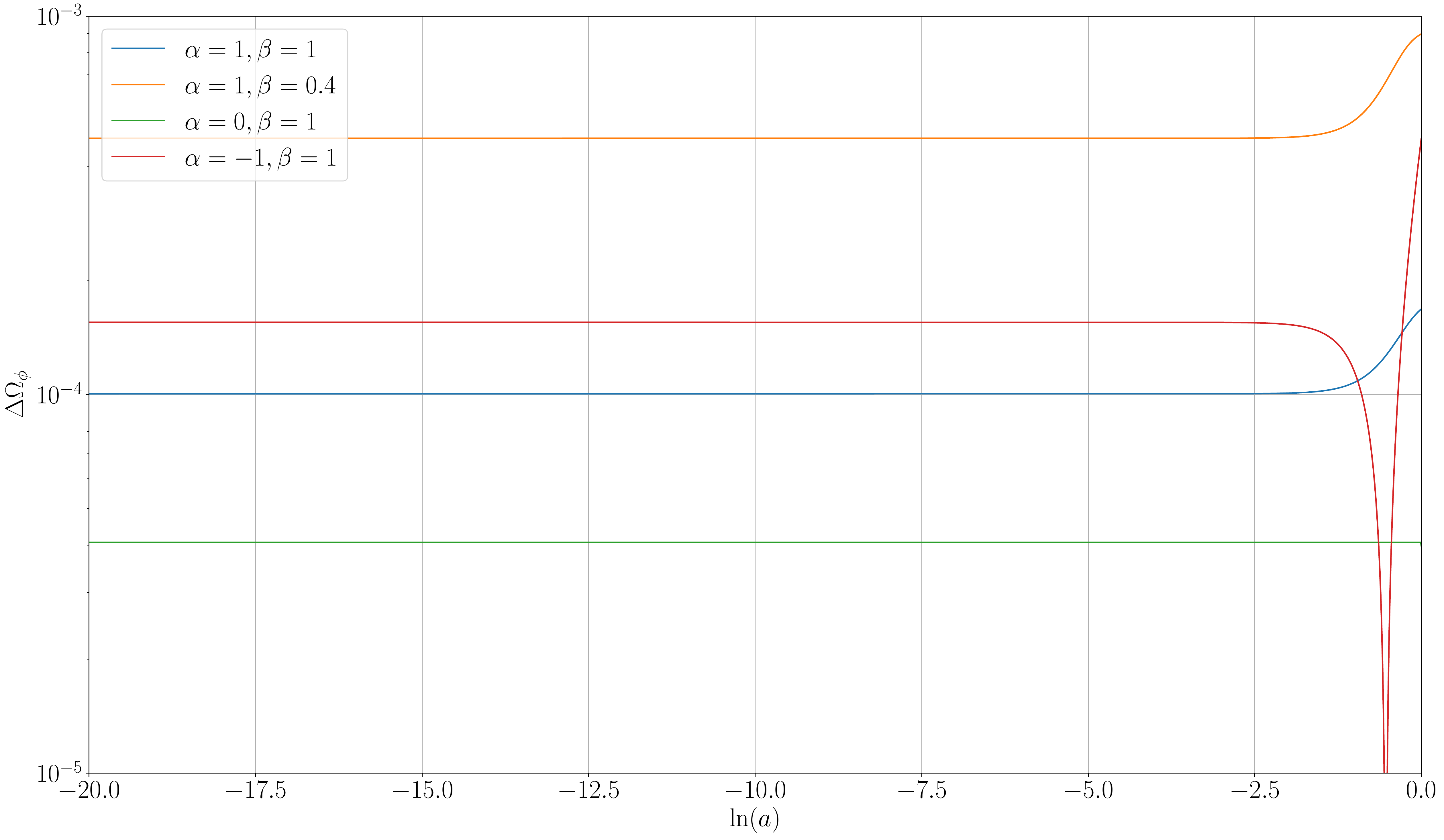}
  \end{center}
  \vspace{-0.5cm}
  \caption{Plot of the relative difference between the analytical and the numerical solutions $\Delta \Omega_\phi$ in equation (\ref{eq:CH(aCS)_fractional_density_relative_difference_ana_num}).}
  \label{fig:CH(aCS)_rinaldi_background_confront}
\end{figure}

In this Appendix we will compare our numerical result with the analytical solution with the analytical solution. This comparison is shown in Fig.\  \ref{fig:CH(aCS)_rinaldi_background_confront}, where we compare our numerical solution $\Omega_{\phi}^{\text{num}}$ (the same used for the fractional density plots in Fig.\ s \ref{fig:CH(aCS)_rinaldi_background} and \ref{fig:CH(aCS)_rinaldi_background2}) with the analytical solution for the fractional density $\Omega_{\phi}$ given in equation (\ref{eq:CH(aCS)_fractional_density_analytical}).

\noindent The quantity plotted as a function of time is
\begin{equation}
\Delta \Omega_\phi =\frac{\left|\Omega_{\phi}^{\text{num}}-\Omega_{\phi}\right|}{\Omega_{\phi}^{\text{num}}},
\label{eq:CH(aCS)_fractional_density_relative_difference_ana_num}
\end{equation}
that is the relative difference between the analytical and the numerical solutions. You can see that the relative difference has a maximum $>0.1$\% for one of the cases.

\section{Equivalence of $G_i$ and $\alpha$ functions methods}
\label{sec:equivalence_of_methods}

In this Appendix we will show that using the $G_i$ functions method is equivalent to use the $\alpha$ functions method. In order to show this, we will consider the relative difference between the fractional density computed with the two methods, that is
\begin{equation}
\Delta \Omega_\phi =\frac{\left|\Omega_{\phi}^{G}-\Omega_{\phi}^\alpha\right|}{\Omega_{\phi}^{G}},
\label{eq:CH(aCS)_fractional_density_relative_difference_G_alpha}
\end{equation}
where $\Omega_{\phi}^{G}$ is the scalar field fractional density computed with the $G_i$ functions method and $\Omega_{\phi}^{\alpha}$ with the $\alpha$ functions method. The difference for the parameters considered is $\Delta \Omega_\phi < 2 \times 10^{-3}$.

\begin{figure}[H]
  \begin{center}
    \includegraphics[width=1.0\textwidth]{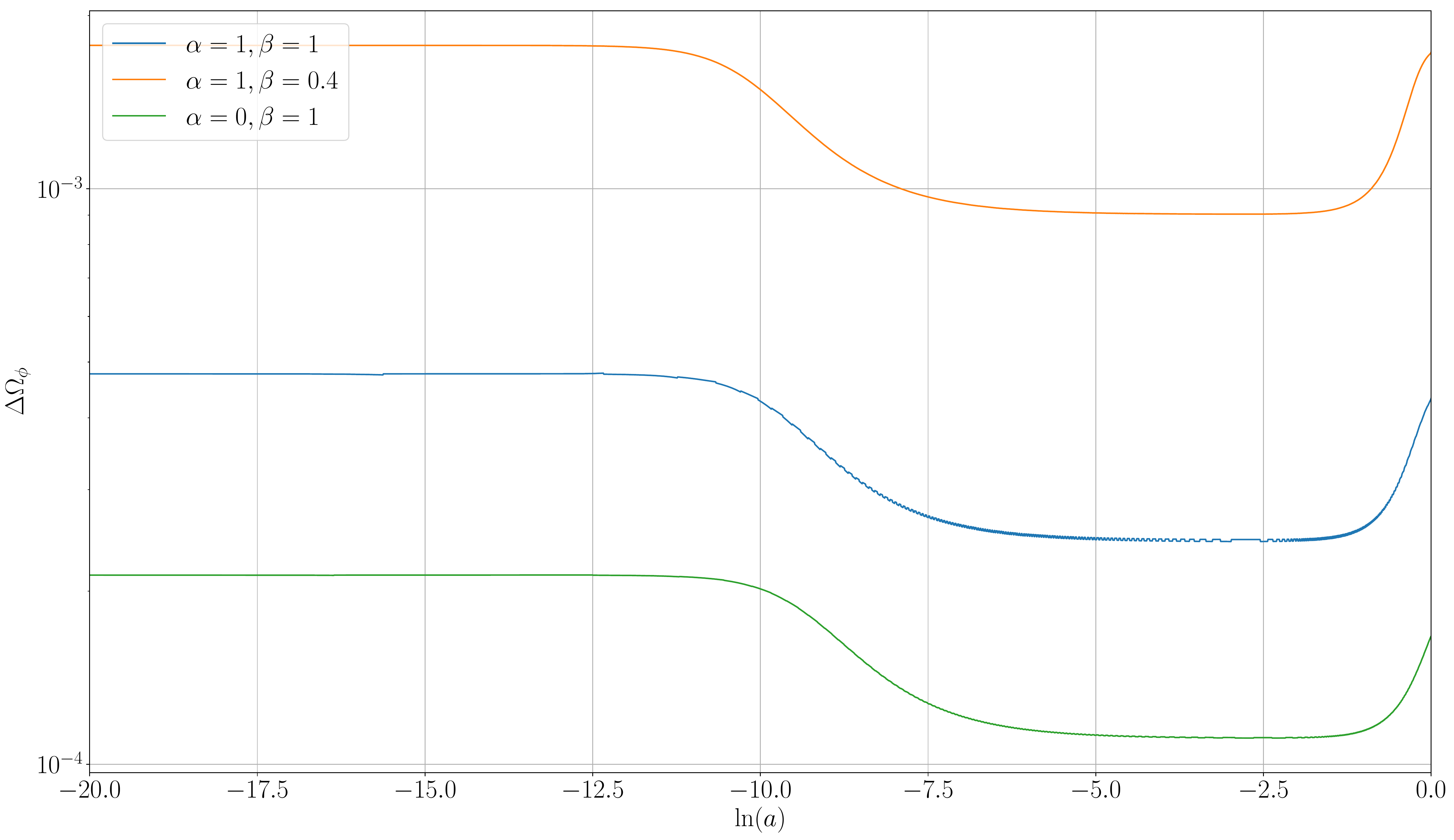}
  \end{center}
  \caption{Plot of the relative difference between the two $G$ and $\alpha$ functions methods (\ref{eq:CH(aCS)_fractional_density_relative_difference_G_alpha}).}
  \label{fig:CH(aCS)_G_alpha_confront}
\end{figure}

\noindent Moreover, we show in Table \ref{tb:CH(aCS)_age_universe_G_alpha} the comparison between the age of the universe $t_0$ computed with the two methods.
\begin{table}[htbp]
\begin{tabular}{ c | c | c | r }
$\alpha$ & $\beta$ & $t_0$ [Gyr] with $G$ & $t_0$ [Gyr] with $\alpha$\\
  \hline			\hline
$1$ & $1$ & $10.603$ & $10.604$\\
$1$ & $0.4$ & $9.710$ & $9.714$\\
$0$ & $1$ & $11.807$ & $11.807$\\ 
\hline
\end{tabular}
\centering
\caption{Table of the results for the age of the universe for the two $G$ and $\alpha$ functions methods.}\label{tb:CH(aCS)_age_universe_G_alpha}
\end{table}

\section{Equivalent form of $G_i$ functions}

In this Appendix we will present an alternative way to write the $G_i$ functions, that is equivalent to (\ref{Gii}), at least at first order perturbation theory. Indeed at first order in perturbation theory, we note that the set of $G_i$ functions found is equivalent to the following set
\begin{equation}\label{eq:Gi_equiv}
G_2 [\phi, X] =-\Lambda+{1\over 2\kappa} \alpha X\,,\quad G_3 [\phi, X] = 0\,,\quad G_4 [\phi, X] = \frac{1}{2} + {1\over 4\kappa} \xi X \,,\quad G_5 [\phi, X] = 0\,.
\end{equation}
This equivalence can be easily checked noticing that the $\alpha$ functions, that govern linear perturbations, computed with the $G_i$ functions (\ref{Gii}) and (\ref{eq:Gi_equiv}) are the same.

The equivalence can also be seen at background level plotting the fractional densities: we found exactly the same result as in Fig.\  \ref{fig:CH(aCS)_rinaldi_background} and \ref{fig:CH(aCS)_rinaldi_background2}.

\section{Additional plots}

\begin{figure}[H]
  \begin{center}
    \includegraphics[width=1\textwidth]{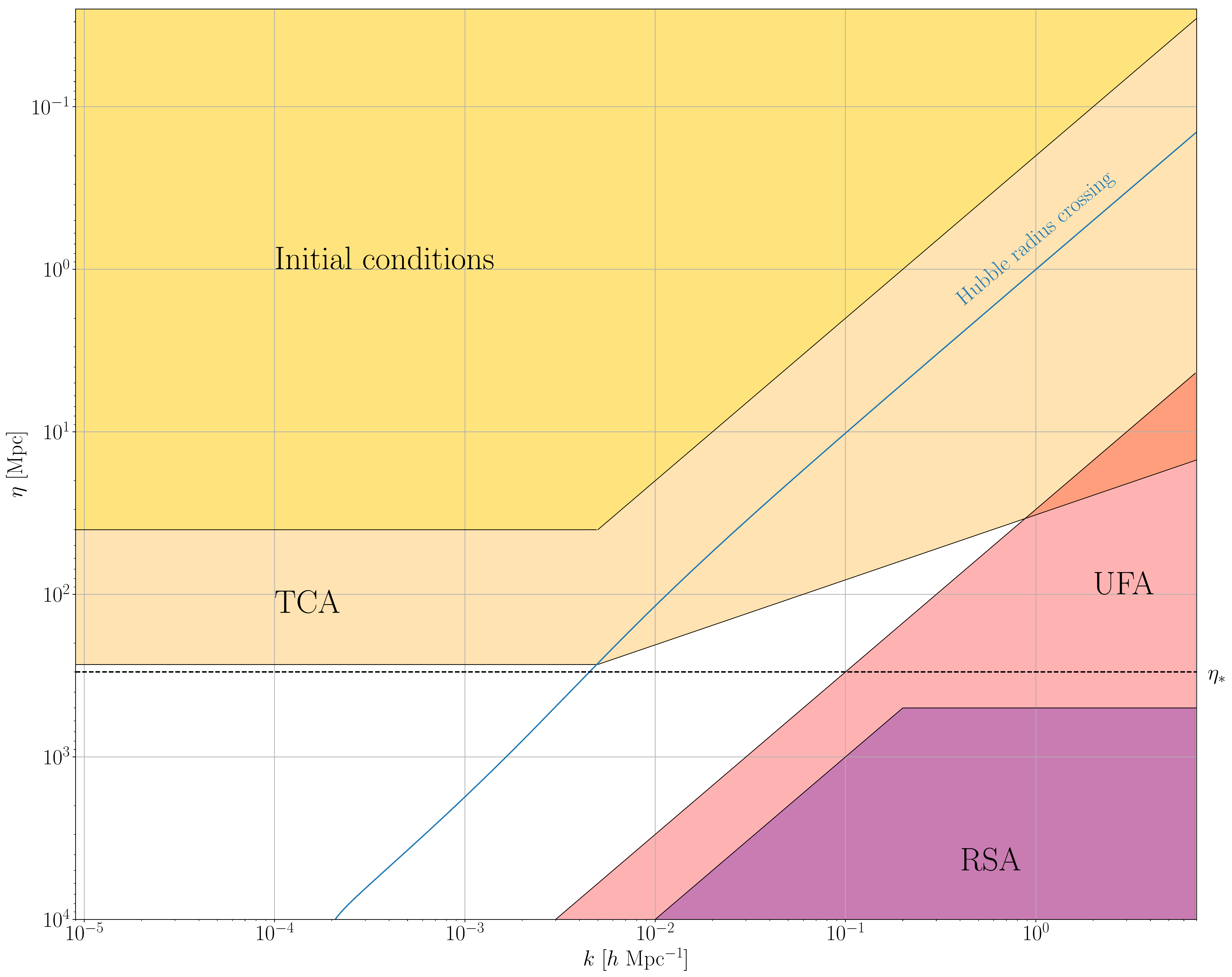}
  \end{center}
  \caption{Plot of the default approximations scheme of (\code{hi\_})\code{CLASS}. This is the same plot shown in the \code{CLASS} article \cite{CLASS}. Every mode $k$ is evaluated in conformal time $\eta$ independently from the others, and each of them has different times where the approximations are switched on and off. Each color represents a different approximation (for more informations, see the \code{CLASS} article). The yellow region is the regime where the initial conditions are considered and the modes are governed by the initial conditions coming from the inflation; the numerical computation starts at the borders of the yellow region using the initial conditions. The white region is where no approximation is used.}
  \label{fig:CH(soft)_approx_class}
\end{figure}

\begin{figure}[H]
\vspace{-0.5cm}
  \begin{center}
    \includegraphics[width=1.0\textwidth]{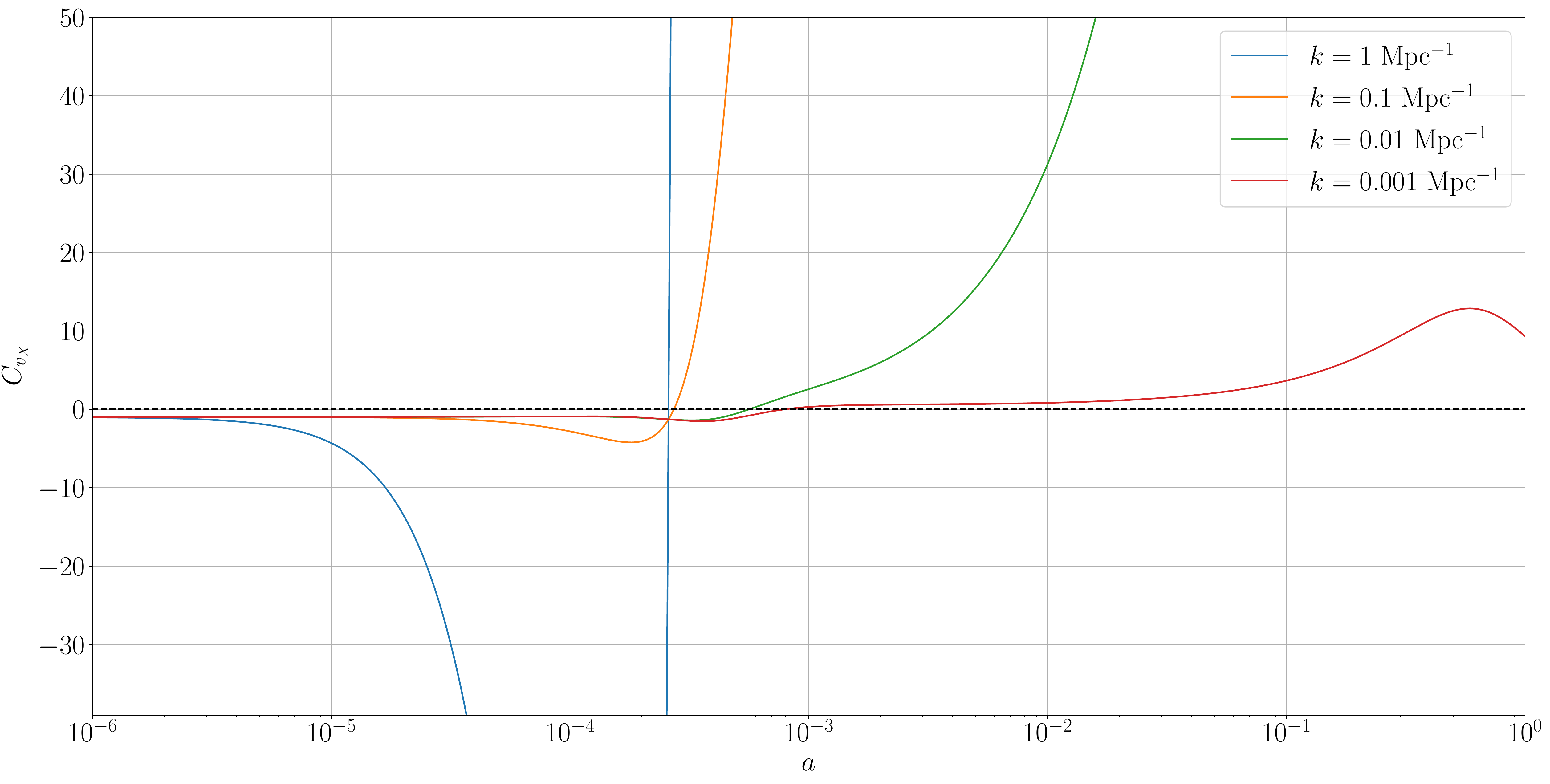}
  \end{center}
  \caption{Plot of $C_{v_X}$ in the case \ref{list:CH(aCS)_enumerate_case2} with $\alpha=1$, $\beta=10^{-10}$ and $\Omega_{cdm}$ equal to the one of $\Lambda$CDM. Above the black dashed line, $C_{v_X}>0$ and the scalar field perturbations differential equation (\ref{eq:CH(aCS)_perturbation_differential_equation}) has an oscillatory behaviour. In this case $C_{v_X}$ does not become negative again as it does in case \ref{list:CH(aCS)_enumerate_case1} in Fig.\  \ref{fig:CH(aCS)_difference_plot2}. Moreover, the only $k$ having a large instability is a very small scale ($k=1$ Mpc$^{-1}$), and this pushes the instability of perturbation functions at large $k$ with respect to the case \ref{list:CH(aCS)_enumerate_case1}, as we see the matter power spectrum in Fig.\  \ref{fig:CH(aCS)_Pk_small_change}.}
  \label{fig:CH(aCS)_difference_plot_case2}
\end{figure}
\begin{figure}[H]
  \begin{center}
    \includegraphics[width=1.0\textwidth]{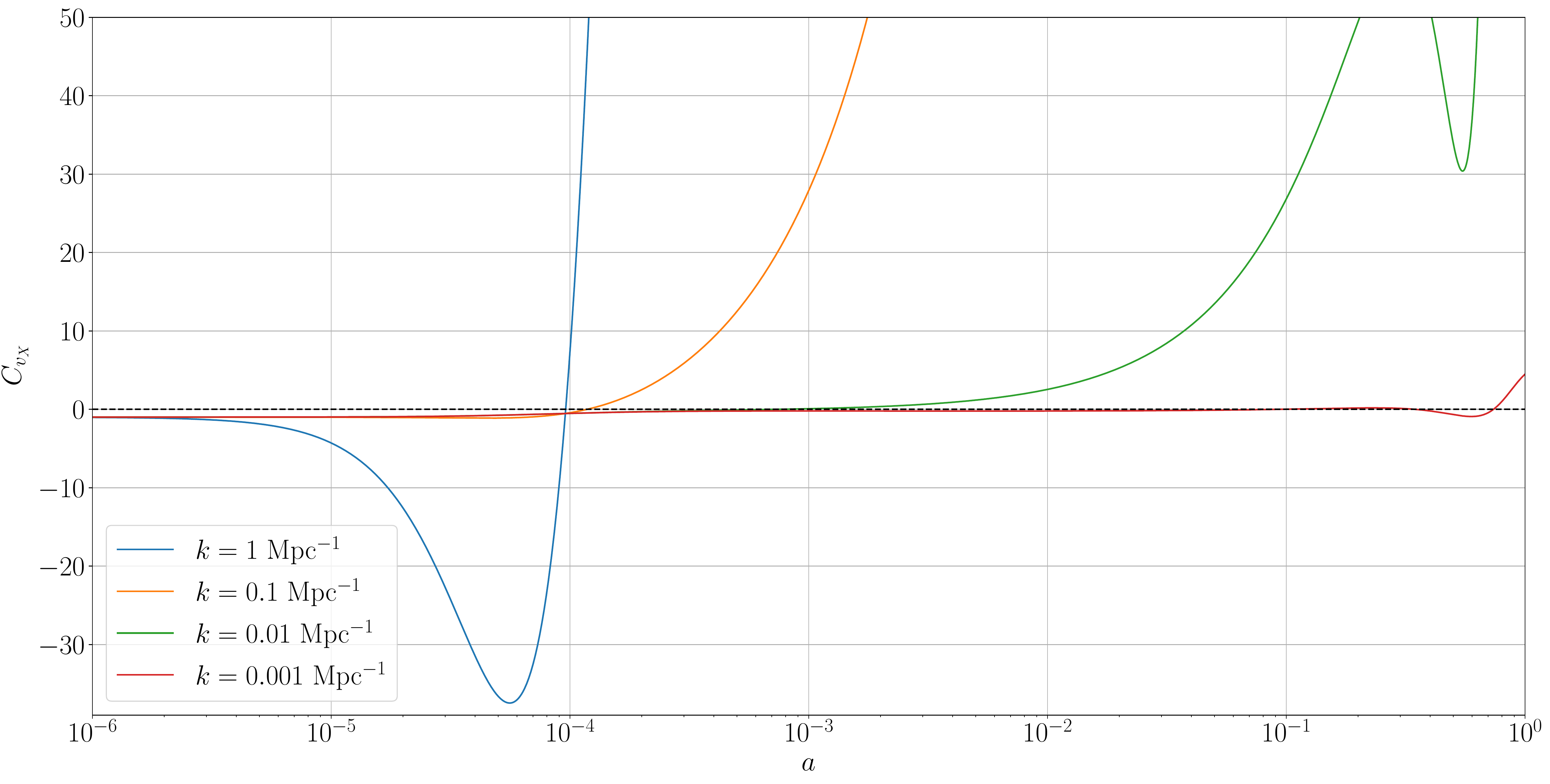}
  \end{center}
  \caption{Plot of $C_{v_X}$ in the case \ref{list:CH(aCS)_enumerate_case3} with $\alpha=1$, $\beta=1$ and $\Omega_{cdm}=0.1$. Above the black dashed line, $C_{v_X}>0$ and the scalar field perturbations differential equation (\ref{eq:CH(aCS)_perturbation_differential_equation}) has an oscillatory behaviour. This case is similar to the first \ref{list:CH(aCS)_enumerate_case1}, but more scales have less instabilities, pushing the instability of perturbation functions at larger $k$ with respect to the first case, as we see the matter power spectrum in Fig.\  \ref{fig:CH(aCS)_Pk_part_cdm}.}
  \label{fig:CH(aCS)_difference_plot_case3}
\end{figure}


\begin{thebibliography}{}

\bibitem{TheLIGOScientific:2017qsa}
  B.~P.~Abbott {\it et al.} [LIGO Scientific and Virgo Collaborations],
  Phys.\ Rev.\ Lett.\  {\bf 119} (2017) no.16,  161101
  
\bibitem{Hou:2017cjy}
  S.~Hou and Y.~Gong,
  Eur.\ Phys.\ J.\ C {\bf 78} (2018) no.3,  247
  
\bibitem{Mukherjee:2017fqz}
  S.~Mukherjee and S.~Chakraborty,
  Phys.\ Rev.\ D {\bf 97} (2018) no.12,  124007
  
\bibitem{Bhattacharya:2016naa}
  S.~Bhattacharya and S.~Chakraborty,
  Phys.\ Rev.\ D {\bf 95} (2017) no.4,  044037

\bibitem{Lombriser:2015sxa}
  L.~Lombriser and A.~Taylor,
  JCAP {\bf 1603} (2016) no.03,  031
  
\bibitem{Lombriser:2016yzn}
  L.~Lombriser and N.~A.~Lima,
  Phys.\ Lett.\ B {\bf 765} (2017) 382
  

\bibitem{Horndeski:1974wa}
  G.~W.~Horndeski,
  Int.\ J.\ Theor.\ Phys.\  {\bf 10} (1974) 363.
  
\bibitem{Creminelli:2017sry}
  P.~Creminelli and F.~Vernizzi,
  Phys.\ Rev.\ Lett.\  {\bf 119} (2017) no.25,  251302

\bibitem{Sakstein:2017xjx}
  J.~Sakstein and B.~Jain,
  Phys.\ Rev.\ Lett.\  {\bf 119} (2017) no.25,  251303
  
\bibitem{Ezquiaga:2017ekz}
  J.~M.~Ezquiaga and M.~Zumalacárregui,
  Phys.\ Rev.\ Lett.\  {\bf 119} (2017) no.25,  251304

\bibitem{Baker:2017hug}
  T.~Baker, E.~Bellini, P.~G.~Ferreira, M.~Lagos, J.~Noller and I.~Sawicki,
  Phys.\ Rev.\ Lett.\  {\bf 119} (2017) no.25,  251301
  
  \bibitem{Rinaldi_2016}  M.~Rinaldi,
  Phys.\ Dark Univ.\  {\bf 16} (2017) 14
  
  \bibitem{amendola}
L.~Amendola,
  Phys.\ Lett.\ B {\bf 301} (1993) 175
 
 \bibitem{maxBH}
 M.~Rinaldi,
  Phys.\ Rev.\ D {\bf 86} (2012) 084048


\bibitem{NSHorn}
A.~Cisterna, T.~Delsate and M.~Rinaldi,
  Phys.\ Rev.\ D {\bf 92} (2015) no.4,  044050
\bibitem{NSHorn2} A.~Cisterna, T.~Delsate, L.~Ducobu and M.~Rinaldi,
  Phys.\ Rev.\ D {\bf 93} (2016) no.8,  084046

 \bibitem{babichev}
 E.~Babichev and C.~Charmousis,
  JHEP {\bf 1408} (2014) 106
%

\bibitem{Zumalacarregui:2016pph}
  M.~Zumalacàrregui, E.~Bellini, I.~Sawicki, J.~Lesgourgues and P.~G.~Ferreira,
  JCAP {\bf 1708} (2017) no.08,  019
  
\bibitem{Diez-Tejedor:2018fue}
  A.~Diez-Tejedor, F.~Flores and G.~Niz,
  Phys.\ Rev.\ D {\bf 97} (2018) no.12,  123524

\bibitem{Planck_2015}
  P.~A.~R.~Ade {\it et al.} [Planck Collaboration],
  Astron.\ Astrophys.\  {\bf 594} (2016) A13

\bibitem{CLASS}
  D.~Blas, J.~Lesgourgues and T.~Tram,
  JCAP {\bf 1107} (2011) 034

\bibitem{tsuji}   A.~De Felice and S.~Tsujikawa,
  JCAP {\bf 1202} (2012) 007
%
%

\bibitem{deRham:2018red}
  C.~de Rham and S.~Melville,
  ``Gravitational Rainbows: LIGO and Dark Energy at its Cutoff,''
  arXiv:1806.09417 [hep-th].
  
\bibitem{Battye:2018ssx}
  R.~A.~Battye, F.~Pace and D.~Trinh,
  Phys.\ Rev.\ D {\bf 98} (2018) no.2,  023504


 \end{thebibliography}
\end{document}